\pdfoutput=1

\documentclass[11pt,twoside,a4paper,cmspaper,final,collab]{cms-tdr}
\def\svnVersion{0249f7f-D}\def\svnDate{2020/11/06}\def\cmsCernNoTag{CERN-EP-2020-143}\def\cmsCernDate{\today}\def\cmsMessage{"Published in Physics Letters B as \href{http://dx.doi.org/10.1016/j.physletb.2020.135988}{\doi{10.1016/j.physletb.2020.135988}.}"}

\begin{document}\cmsNoteHeader{SMP-19-008}

\hyphenation{had-ron-i-za-tion}
\hyphenation{cal-or-i-me-ter}
\hyphenation{de-vices}
\newlength\cmsTabSkip\setlength{\cmsTabSkip}{1ex}
\newlength\cmsFigWidth
\ifthenelse{\boolean{cms@external}}{\setlength\cmsFigWidth{0.49\textwidth}}{\setlength\cmsFigWidth{0.65\textwidth}}
\newlength\cmsFigWidthSplit
\ifthenelse{\boolean{cms@external}}{\setlength\cmsFigWidthSplit{0.49\textwidth}}{\setlength\cmsFigWidthSplit{0.49\textwidth}}
\ifthenelse{\boolean{cms@external}}{\providecommand{\cmsLeft}{upper\xspace}}{\providecommand{\cmsLeft}{left\xspace}}
\ifthenelse{\boolean{cms@external}}{\providecommand{\cmsRight}{lower\xspace}}{\providecommand{\cmsRight}{right\xspace}}

\ifthenelse{\boolean{cms@external}}{\providecommand{\cmsULeft}{upper left\xspace}}{\providecommand{\cmsULeft}{left\xspace}}
\ifthenelse{\boolean{cms@external}}{\providecommand{\cmsURight}{upper right\xspace}}{\providecommand{\cmsURight}{second from left\xspace}}
\ifthenelse{\boolean{cms@external}}{\providecommand{\cmsLLeft}{lower left\xspace}}{\providecommand{\cmsLLeft}{third from left\xspace}}
\ifthenelse{\boolean{cms@external}}{\providecommand{\cmsLRight}{lower right\xspace}}{\providecommand{\cmsLRight}{right\xspace}}

\newcommand {\sdev}{\unit{SD}}
\newcommand{\MJ}{\ensuremath{m_\text{J}^\text{SD}}\xspace}
\newcommand{\Jg}{\ensuremath{\text{J}\Pgg}\xspace}
\newcommand{\Hg}{\ensuremath{\PH\Pgg}\xspace}
\newcommand{\mll}{\ensuremath{m_{\Pell\Pell}}}
\newcommand{\dyll}{\ensuremath{\PZ/\Pgg^*\to\Pell^+\Pell^-}}
\newcommand{\jj}{\ensuremath{\text{jj}}\xspace}
\newcommand{\alphahat}{\ensuremath{\hat{\alpha}}\xspace}
\newcommand{\loperator}[2]{\ensuremath{\mathcal{L}_{\text{#1,#2}}}\xspace}
\newcommand{\fcoeff}[2]{\ensuremath{f_{\text{#1,#2}}}\xspace}
\newcommand{\fcoefflam}[2]{\ensuremath{f_{\text{#1,#2}}/\Lambda^{4}}\xspace}
\newcommand{\ubound}{\ensuremath{U_{\text{bound}}}\xspace}
\newcommand{\escat}{\ensuremath{E_{\text{scattering}}}\xspace}
\newcommand{\dnll}{\ensuremath{\Delta\text{NLL}}\xspace}
\newcommand{\dejj}{\ensuremath{\Delta\etajj}\xspace}
\newcommand{\dphilmet}{\ensuremath{\Delta\phi_{\Pell,\ptmiss}}\xspace}
\newcommand{\muhat}{\ensuremath{\hat{\mu}}\xspace}
\newcommand{\fremu}{\ensuremath{f_{\Pe{}(\Pgm{})}}\xspace}
\newcommand{\frlep}{\ensuremath{f_{\Pell}}\xspace}
\newcommand{\muR}{\ensuremath{\mu_{\text{R}}}\xspace}
\newcommand{\muF}{\ensuremath{\mu_{\text{F}}}\xspace}
\newcommand{\thetab}{\ensuremath{\boldsymbol{\theta}}\xspace}
\newcommand{\thetabhat}{\ensuremath{\hat{\thetab}}\xspace}
\newcommand{\thetabdhat}{\ensuremath{\hat{\thetabhat}}\xspace}
\newcommand{\sigmag}{\ensuremath{\sigma_\text{g}}\xspace}
\newcommand{\sigmagew}{\ensuremath{\sigmag^\text{EW}}\xspace}
\newcommand{\sigmagqcd}{\ensuremath{\sigmag^\text{QCD}}\xspace}
\newcommand{\sigmaee}{\ensuremath{\sigma_{\eta\eta}}\xspace}
\newcommand{\sigfid}{\ensuremath{\sigma^{\text{fid}}}\xspace}
\newcommand{\sigfidew}{\ensuremath{\sigfid_{\text{EW}}}\xspace}
\newcommand{\sigfidewq}{\ensuremath{\sigfid_{\text{EW+QCD}}}\xspace}
\newcommand{\accgf}{\ensuremath{\text{a}_\text{gf}}\xspace}
\newcommand{\alphagf}{\ensuremath{\alpha_\text{gf}}\xspace}
\newcommand{\alphaewgf}{\ensuremath{\alphagf^\text{EW}}\xspace}
\newcommand{\alphaqcdgf}{\ensuremath{\alphagf^\text{QCD}}\xspace}
\newcommand{\alphatest}{\ensuremath{\alpha_\text{test}}\xspace}
\newcommand{\WWjj}{\ensuremath{\PW^{\pm}\PW^{\pm}\jj}\xspace}
\newcommand{\ttg}{\ensuremath{\ttbar\Pgg}\xspace}
\newcommand{\WW}{\ensuremath{\PW{}\PW}\xspace}
\newcommand{\WZ}{\ensuremath{\PW{}\PZ}\xspace}
\newcommand{\ZZ}{\ensuremath{\PZ{}\PZ}\xspace}
\newcommand{\Wg}{\ensuremath{\PW{}\Pgg}\xspace}
\newcommand{\VV}{\ensuremath{\PV{}\PV}\xspace}
\newcommand{\pp}{\ensuremath{\Pp{}\Pp}\xspace}
\newcommand{\mWg}{\ensuremath{m_{\Wg{}}}\xspace}
\newcommand{\Wgjj}{\ensuremath{\PW{}\Pgg{}\jj}\xspace}
\newcommand{\WWgg}{\ensuremath{\PW{}\PW{}\Pgg{}\Pgg}\xspace}
\newcommand{\Zg}{\ensuremath{\PZ{}\Pgg}\xspace}
\newcommand{\gpjets}{\Pgg{}+jets\xspace}
\newcommand{\Wpjets}{\PW{}+jets\xspace}
\newcommand{\Wgpjets}{\PW{}\Pgg{}+jets\xspace}
\newcommand{\Zpjets}{\PZ{}+jets\xspace}
\newcommand{\Zgpjets}{\PZ{}\Pgg{}+jets\xspace}
\newcommand{\lepg}{\ensuremath{\Pell{}\Pgg}\xspace}
\newcommand{\mjj}{\ensuremath{m_{\jj}}\xspace}
\newcommand{\mlepg}{\ensuremath{m_{\lepg}}\xspace}
\newcommand{\lnugjj}{\ensuremath{\Pell{}\Pgn{}\Pgg{}\jj}\xspace}
\newcommand{\mTW}{\ensuremath{\mT^\PW}\xspace}
\newcommand{\pz}{\ensuremath{p_{\text{z}}}\xspace}
\newcommand{\ptX}[1]{\ensuremath{\pt^{#1}}\xspace}
\newcommand{\reliso}{\ensuremath{\text{Iso}}\xspace}
\newcommand{\PUpt}{\ensuremath{\ptX{\text{PU}}}\xspace}
\newcommand{\ptg}{\ensuremath{\ptX{\Pgg}}\xspace}
\newcommand{\ptlep}{\ensuremath{\ptX{\Pell}}\xspace}
\newcommand{\ptjj}{\ensuremath{\ptX{\jj}}\xspace}
\newcommand{\ptj}{\ensuremath{\ptX{\text{j}}}\xspace}
\newcommand{\ptW}{\ensuremath{\ptX{\PW}}\xspace}
\newcommand{\ptZ}{\ensuremath{\ptX{\PZ}}\xspace}
\newcommand{\etaX}[1]{\ensuremath{\eta_{#1}}\xspace}
\newcommand{\etaSC}{\etaX{\text{SC}}}
\newcommand{\etag}{\etaX{\Pgg{}}}
\newcommand{\etajj}{\etaX{\jj}}
\newcommand{\etaj}{\etaX{\text{j}}}
\newcommand{\etaW}{\etaX{\PW{}}}
\newcommand{\etaZ}{\etaX{\PZ{}}}
\newcommand{\etalep}{\etaX{\Pell{}}}
\newcommand{\ptch}{\ensuremath{\ptX{\text{charged}}}\xspace}
\newcommand{\ptneu}{\ensuremath{\ptX{\text{neutral}}}\xspace}
\newcommand{\sumg}{\ensuremath{\sum\ptg}\xspace}
\newcommand{\sumch}{\ensuremath{\sum\ptch}\xspace}
\newcommand{\sumneu}{\ensuremath{\sum\ptneu}\xspace}
\newcommand{\phiWg}{\ensuremath{\phi_{\Wg{}}}\xspace}
\newcommand{\phiDijet}{\ensuremath{\phi_{\text{j1,j2}}}\xspace}
\newcommand{\kinsysY}{\ensuremath{\abs{y_{\Wg{}}-(y_{\text{j1}}+y_{\text{j2}})/2}}\xspace}
\newcommand{\kinsysPhi}{\ensuremath{\abs{\phiWg - \phiDijet}}\xspace}
\newcommand{\eletogam}{\ensuremath{\Pe\to\Pgg}\xspace}
\providecommand{\cmsTable}[1]{\resizebox{\textwidth}{!}{#1}}

\cmsNoteHeader{SMP-19-008} 
\title{Observation of electroweak production of \texorpdfstring{\Wg}{W gamma} with two jets in proton-proton collisions at \texorpdfstring{$\sqrt{s}=13\TeV$}{sqrt(s)=13 TeV}}

\date{\today}

\abstract{
A first observation is presented for the electroweak production of a \PW~boson, a photon, and two jets in proton-proton collisions. The \PW~boson decays are selected by requiring one identified electron or muon and an imbalance in transverse momentum. The two jets are required to have a high dijet mass and a large separation in pseudorapidity. The measurement is based on data collected with the CMS detector at a center-of-mass energy of 13\TeV, corresponding to an integrated luminosity of 35.9\fbinv . The observed (expected) significance for this process is 4.9 (4.6) standard deviations. After combining with previously reported CMS results at 8\TeV, the observed (expected) significance is 5.3 (4.8) standard deviations. The cross section for the electroweak \Wgjj production in a restricted fiducial region is measured as $20.4\pm 4.5\unit{fb}$ and the total cross section for \Wg production in association with 2 jets in the same fiducial region is $108\pm 16\unit{fb}$. All results are in good agreement with recent theoretical predictions. Constraints are placed on anomalous quartic gauge couplings in terms of dimension-8 effective field theory operators.
}

\hypersetup{
pdfauthor={CMS Collaboration},
pdftitle={Observation of the electroweak production of W gamma in association with two jets in proton-proton collisions at sqrt(s) = 13 TeV},
pdfsubject={CMS},
pdfkeywords={CMS, physics, vector boson scattering}}

\maketitle
\section{Introduction}
\label{sec:introduction}
After the discovery of the Higgs boson at the CERN LHC~\cite{Aad:2012tfa,Chatrchyan:2012ufa,Chatrchyan_2013}, one of the primary goals of high-energy physics is to examine the details of the mechanism of electroweak (EW) symmetry breaking, e.g., through measurements of the properties of the Higgs boson. 
Vector boson scattering (VBS) processes comprise an independent and complementary method to study EW symmetry breaking. 
The nonabelian nature of gauge interactions in the standard model (SM) leads to a rich variety of VBS processes with unique features and opportunities to probe physics beyond the SM (BSM).

The high energy and luminosity of the LHC make it possible to study the rare VBS processes in detail. 
The CMS Collaboration reported the EW production of two \PW~bosons of same electric charge produced in association with two jets (\WWjj), with a significance of 5.5 standard deviations (SD) based on the initial proton-proton (\pp) data collected at 13\TeV~\cite{Sirunyan:2017ret}. 
There have been additional VBS results from both the ATLAS and CMS Collaborations. 
Notably, ATLAS observed EW (\WWjj) production with a significance of 6.5\sdev~\cite{Aaboud_2019}. 
CMS recently reported an observation of \WZ VBS events at a significance of 6.8\sdev~\cite{SMP-19-012}, along with further studies in the \WWjj channel, based on data collected at 13\TeV. 
Moreover, VBS processes involving a photon in the final state, \Wg and \Zg scattering, were also reported by ATLAS and CMS, based on data collected at $\sqrt{s}=8\TeV$, corresponding to an integrated luminosity of approximately 20\fbinv~\cite{Aaboud:2017pds,Khachatryan:2017jub,Khachatryan:2016vif}. 
The observed (expected) significance for \Wg scattering from CMS was 2.7 (1.5)\sdev. 
For \Zg scattering ATLAS and CMS observed (expected) significances of 2.0 (1.8) and 3.0 (2.1)\sdev, respectively, based on the SM prediction. 
A recent update on \Zg scattering from CMS, based on the initial data collected at 13\TeV combined with 8\TeV results~\cite{Sirunyan:2020tlu}, reported an observed (expected) significance of 4.7 (5.5)\sdev.

\begin{figure*}[ht!]
  \centering
    \includegraphics[height=0.21\textwidth,width=0.23\textwidth]{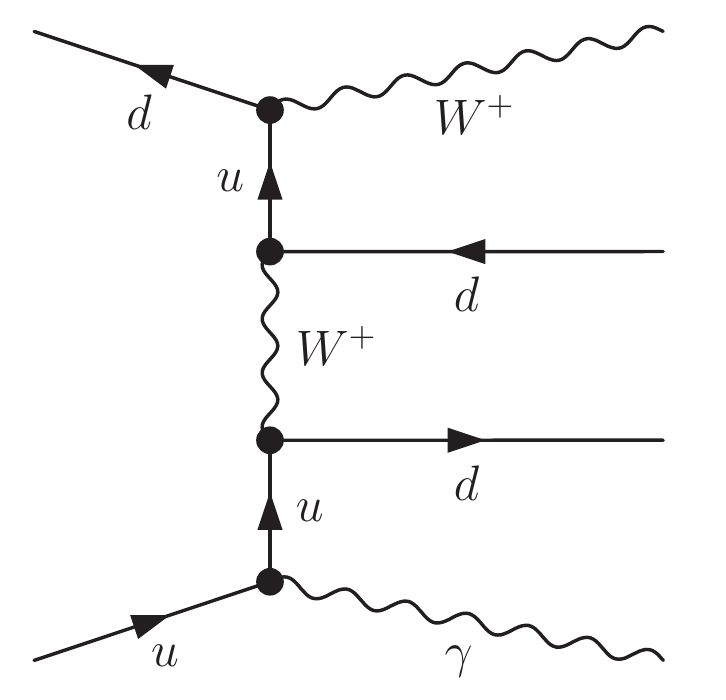}
    \includegraphics[height=0.21\textwidth,width=0.23\textwidth]{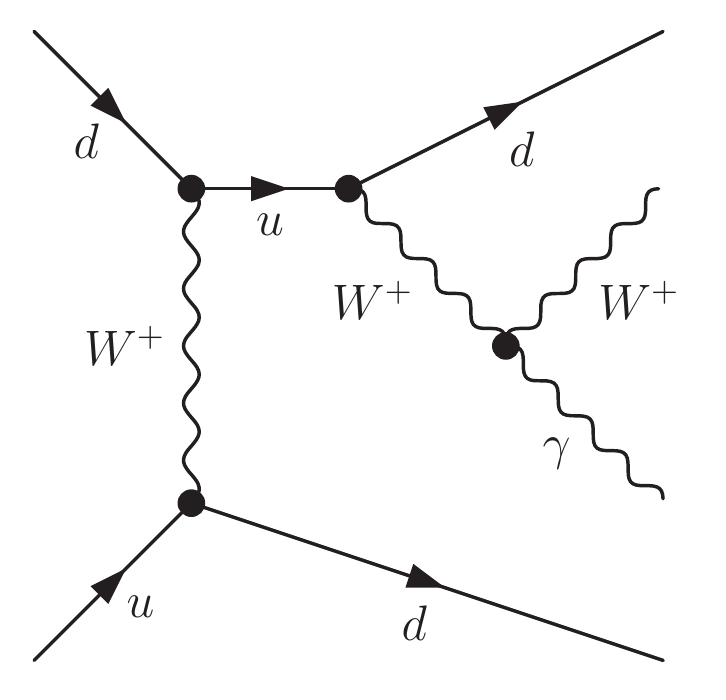}
    \includegraphics[height=0.21\textwidth,width=0.23\textwidth]{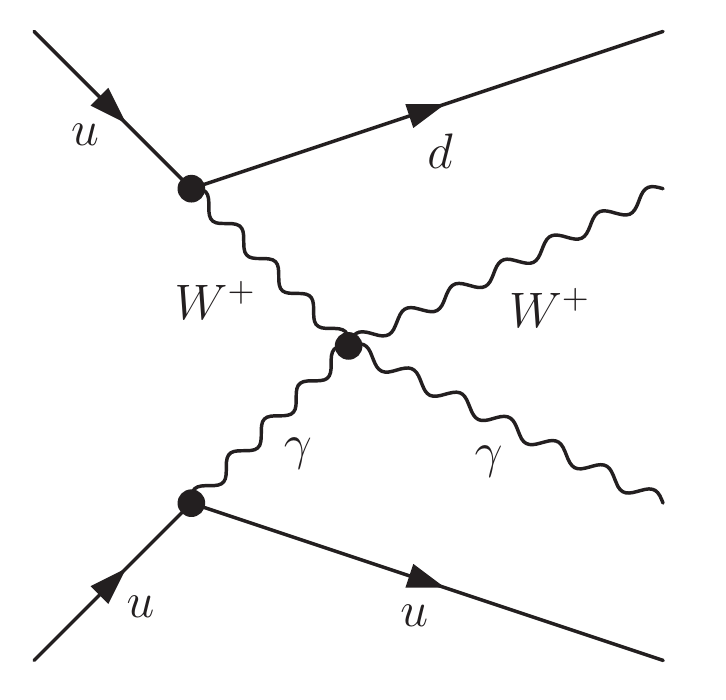}
    \includegraphics[height=0.21\textwidth,width=0.23\textwidth]{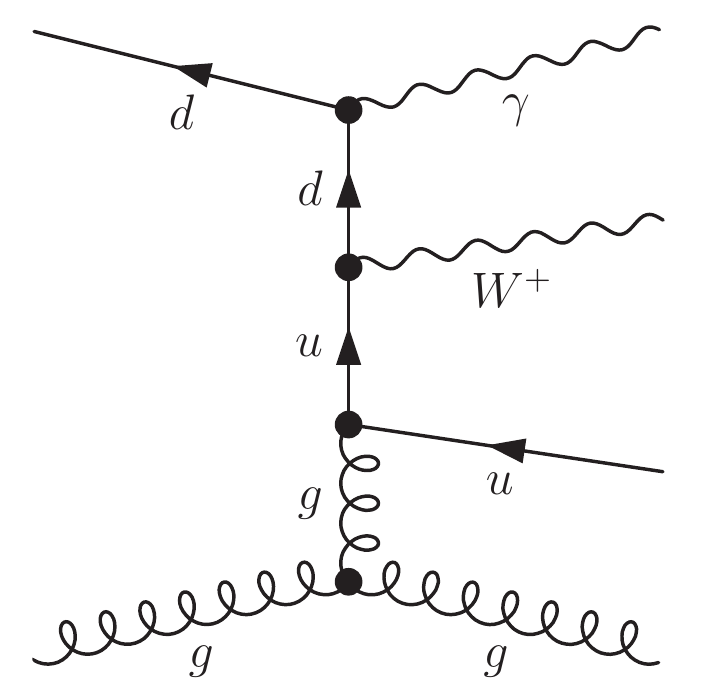}
  \caption{Representative diagrams for \lnugjj production at the LHC for EW production (left), EW production through triple (middle left) and quartic (middle right) gauge boson couplings, and QCD-induced processes (right).}
  \label{fig:wa_feynman}
\end{figure*}

This paper presents a measurement of VBS in the \Wg channel at $\sqrt{s}=13\TeV$. 
As shown in Fig.~\ref{fig:wa_feynman}, the signal process includes both VBS and non-VBS EW diagrams, such as EW contributions through triple and quartic gauge couplings. 
QCD-induced production of \Wgjj can also take place, as shown in the diagram on the right, with both jets originating from QCD vertices. 
The diagrams shown are representative of the many possibilities in the SM. 
The effects of BSM physics, such as anomalous triple and quartic gauge couplings (aTGC and aQGC), are also possible~\cite{paper_aqgc}.
While aTGC are well constrained by other processes including Higgs boson and diboson production, VBS studies are more sensitive to aQGC.

The data correspond to an integrated luminosity of $35.9 \pm 0.9\fbinv$ collected during 2016 using the CMS detector~\cite{Chatrchyan:2008zzk} at the LHC. 
For measuring the EW \Wgjj production, candidate events are selected by requiring one identified lepton (either an electron or muon), one identified photon, two jets with a large rapidity separation and a large dijet invariant mass (\mjj), and a moderate imbalance in transverse momentum, \ptmiss. 
This selection reduces the contribution from the strong (QCD) production of jets produced together with the \PW~boson and the photon, making the experimental signature an ideal topology for VBS \Wg studies. 
The interference among the VBS diagrams ensures the unitarity of the VBS cross section in the SM at high energy, and an interference is also expected between QCD and EW processes~\cite{Khoze:2002fa,PhysRevD.69.093004}.

\section{The CMS detector}
\label{sec:detector}
The central feature of the CMS~\cite{Chatrchyan:2008zzk} apparatus is a superconducting solenoid of 6\unit{m} internal diameter, providing a magnetic field of 3.8\unit{T}.
A silicon pixel and strip tracker, a lead tungstate crystal electromagnetic calorimeter (ECAL), and a brass and scintillator hadron calorimeter (HCAL), each composed of a barrel and two endcap sections reside within the volume of the solenoid.
Forward calorimeters extend the coverage provided by the barrel and endcap detectors up to a pseudorapidity of $\abs{\eta}=5$.
Muons are detected in gas-ionization chambers embedded in the steel flux-return yoke outside the solenoid.

Events of interest are selected using a two-tiered trigger system~\cite{Khachatryan:2016bia}.
The first level (L1), composed of specialized hardware processors, uses information from the calorimeters and muon detectors to select events at a rate of around 100\unit{kHz} with a latency of 4\mus.
The second level consists of a farm of processors running a version of the full event reconstruction software optimized for fast processing that reduces the event rate to around 1\unit{kHz} before data storage.

A more detailed description of the CMS detector, together with a definition of the coordinate system and kinematic variables, can be found in Ref.~\cite{Chatrchyan:2008zzk}.

\section{Signal and background simulation}
\label{sec:simulation}
The signal and background processes are simulated using the Monte Carlo (MC) generator \MGvATNLO (MG5)~\cite{MGatNLO}.
The EW \Wgjj signal is simulated at leading order (LO) using version 2.6.0.
The main background from QCD \Wg is simulated with up to one jet in the matrix element calculation at next-to-leading order (NLO) with version 2.4.2, using the FxFx scheme~\cite{Frederix:2012ps} to merge jets from the matrix element calculation and parton showering.
The interference between the EW and QCD processes is predicted to 1--3\% in the signal region and is treated as a systematic uncertainty.
Other background contributions include diboson \VV processes (\WW, \WZ, \ZZ) simulated at LO with \PYTHIA~8.212~\cite{Sj_strand_2015}, single top quark processes simulated with \POWHEG~2.0~\cite{POWHEG}, and \ttg production simulated at NLO with MG5 using the FxFx jet merging scheme.
Cross sections evaluated at NLO in the QCD coupling strength (\alpS) are used to normalize these simulated event samples.

The \PYTHIA package, with the CUETP8M1~\cite{Skands:2014pea,Khachatryan:2015pea} tune, is used for parton showering, hadronization, and underlying-event simulation.
The NNPDF~3.0 set~\cite{Ball:2014uwa} of parton distribution functions (PDFs) is used as default.
All simulated events are processed through a \GEANTfour~\cite{geant4} simulation of the CMS detector.
Factors determined by a tag-and-probe technique ~\cite{tagandprobe} are used to correct the differences between data and simulation in the trigger efficiency, as well as the reconstruction and identification (ID) efficiencies.
Additional overlapping \pp interactions (pileup) are superimposed over the hard scattering interaction with a distribution of primary vertices matching that obtained from the collision data. The MC samples are analyzed using the same procedures as the data.

\section{Event reconstruction}
\label{sec:reconstruction}
The particle-flow (PF) algorithm~\cite{Sirunyan:2017ulk} reconstructs and identifies each individual particle in an event, through an optimized combination of information from the various elements of the CMS detector.
The energy of photons is obtained from the ECAL measurement.
The energy of electrons is determined from a combination of the electron momentum at the primary interaction vertex as determined in the tracker, the energy of the corresponding ECAL cluster, and the energy sum of all bremsstrahlung photons spatially compatible with originating from the electron track.
The energy of muons is obtained from the curvature of the corresponding track.
The energy of charged hadrons is determined from a combination of their momentum measured in the tracker and the matching ECAL and HCAL energy depositions, corrected for the response of the calorimeters to hadronic showers.
Finally, the energy of neutral hadrons is obtained from the corresponding corrected ECAL and HCAL energies.
The PF candidates are used for a variety of purposes in this analysis, such as evaluating electron, muon, and photon isolation variables, reconstructing jets, and computing the \ptmiss in the event, as described below.

The reconstructed vertex with the largest value of summed jet $\pt^2$ is taken to be the primary \pp interaction vertex~\cite{TRK-11-001}.
The jets are clustered using the anti-\kt jet finding algorithm~\cite{Cacciari:antikt,Cacciari:fastjet1} using the tracks assigned to candidate vertices as inputs.

Electron candidates used in the selection of events for this analysis are reconstructed within $\abs{\eta}<2.5$ for $\pt>25\GeV$.
The electrons are also required to pass additional identification criteria: selection on the relative amount of energy deposited in the HCAL, a match of the trajectory in the tracker with the position of the ECAL cluster~\cite{Khachatryan:2015hwa}, the number of missing measurements in the tracker, the compatibility of the electron to originate from the primary vertex, and \sigmaee, a parameter that quantifies the spread in $\eta$ of the shower in the ECAL, as discussed in Section~\ref{sec:bkg_estimation}.
Electrons identified as arising from photon conversions are rejected~\cite{Khachatryan:2015hwa,electron_7tev}.
A high-quality ID selection is used to identify electrons in the final state, and a loose selection is used to identify electrons for vetoing events containing additional leptons.

Muons are reconstructed from information in the muon system and the tracker within $\abs{\eta}<2.4$ and $\pt>20\GeV$~\cite{Sirunyan_2018}.
Muon candidates must satisfy ID criteria based on the number of measurements in the muon system and the tracker, the number of matched muon-detector planes, the quality of the combined fit to the track, and the compatibility of the muon to originate from the primary vertex.
A high-quality ID~\cite{Sirunyan_2018} is used to identify muons in the final state, and a loose ID~\cite{Sirunyan_2018} is used to identify muons for vetoing events with additional leptons.

Another selection on an isolation variable (\reliso) is applied for both electrons and muons.
Iso is defined relative to the lepton \pt by summing the \pt of the charged hadrons and neutral particles in geometrical cones of $\DR=\sqrt{\smash[b]{(\Delta\eta)^2+(\Delta\phi)^2}}=0.3\,(0.4)$ around the electron (muon) trajectory:
\begin{linenomath}
\ifthenelse{\boolean{cms@external}}{
\begin{multline*}
I^{\ell} = \Bigg( \sum  \pt^\text{charged}\\
+ \text{max}\Big[ 0, \sum \pt^\text{neutral}
+  \sum \pt^{\gamma} - \pt^{\mathrm{PU}}  \Big] \Bigg) \Big/  \pt^{\ell}.
\end{multline*}
}{
\begin{equation*}
I^{\ell} = \left( \sum  \pt^\text{charged} + \text{max}\left[ 0, \sum \pt^\text{neutral} + \sum \pt^{\gamma} - \pt^{\mathrm{PU}}  \right] \right) /  \pt^{\ell}.
\end{equation*}
}
\end{linenomath}
where \sumch is the scalar sum of the transverse momenta of charged hadrons originating from the primary vertex, and \sumneu and \sumg are, respectively, the scalar \pt sums of neutral hadrons and photons.
To mitigate pileup (PU) effects, only charged hadrons originating at the primary vertex are included.
For the neutral-hadron and photon components, an estimate of the expected PU contribution (\PUpt)~\cite{Sirunyan_2020} is subtracted.
For electrons, \PUpt is evaluated using the ``jet area'' method described in Ref.~\cite{jetarea_method}, whereas for muons, \PUpt is assumed to be one half of the scalar \pt sum deposited in the isolation cone by charged particles not associated with the primary vertex.
The factor of one half corresponds to the approximate ratio of neutral to charged hadrons produced in the hadronization of PU interactions.
Electrons passing the high-quality (loose) ID selection are considered isolated if $\reliso<0.0695\,(0.175)$ if the pseudorapidity (\etaSC) of the ECAL cluster is $\abs{\etaSC}<1.479$, or $\reliso<0.0821\,(0.159)$ if $1.479<\abs{\etaSC}<2.5$.
Muons are considered isolated if $\reliso<0.15\,(0.25)$ for the high-quality (loose) ID selection.

Photon reconstruction~\cite{photon_8tev} is similar to that of electrons, and is performed in the region of $\abs{\eta}<2.5$ and for $\pt>20\GeV$, excluding the ECAL transition region of $1.444<\abs{\eta}<1.566$.
To minimize photon misidentification, photon candidates must: pass an electron veto; satisfy criteria based on the distribution of energy deposited in the ECAL and HCAL; satisfy criteria on the isolation variables constructed from the kinematic inputs of the charged and neutral hadrons; and have no other photons near the photon of interest.
A high-quality ID~\cite{photon_8tev} is used to identify prompt photons (\ie, not originating from hadron decays) in the final state, and a loose ID~\cite{photon_8tev} to identify nonprompt photons, which are mainly products of neutral pion decay.

Jets are reconstructed from PF objects using the anti-\kt jet clustering algorithm~\cite{Cacciari:antikt} with a distance parameter of 0.4.
To reduce the contamination from PU, charged PF candidates in the tracker acceptance of $\abs{\eta}<2.4$ are excluded from jet clustering when they are associated with PU vertices~\cite{Sirunyan:2017ulk}.
The contribution from neutral PU particles to the jet energy is corrected based on the projected area of the jet on the front face of the calorimeter~\cite{jetarea_method}.
For this analysis, jets are required to have $\abs{\eta}<4.7$ and $\pt>30\GeV$.
A jet energy correction, similar to the one developed for 8\TeV collisions~\cite{jer}, is obtained from dedicated studies we performed on both data and simulated events (typically involving dijet, {\Pgg}+jet, {\PZ}+jet, and multijet production).
Other residual corrections are applied to the data as functions of \pt and $\eta$ to correct for the small differences between data and simulation.
Additional quality criteria are applied to jet candidates to remove spurious jet-like features originating from isolated noise patterns~\cite{CMS:2010xta} in the calorimeters or the tracker.

The vector \ptvecmiss is computed as the negative of the vector sum of the \pt of all the PF candidates in an event~\cite{Sirunyan:2019kia}, and its magnitude is denoted as \ptmiss.
The jet energy corrections are propagated to the \ptvecmiss.
The data to simulation efficiency ratios are used as scale factors to correct the simulated event yields.

\section{Event selection}
\label{sec:selection}
Candidate events are selected by requiring exactly one electron (muon) with $\pt>30\GeV$ and $\abs{\etalep}<2.5\,(2.4)$, with transverse mass of the \PW~boson $\mTW>30\GeV$.
We define \mTW as $\sqrt{\smash[b]{2\ptlep\ptmiss[1-\cos{(\dphilmet)}]}}$, where \ptlep is the \pt of the lepton and \dphilmet is the azimuthal angle between the lepton and the \ptvecmiss directions.
Events are required to contain a well-identified and isolated photon with $\ptg>25\GeV$, $\ptmiss>30\GeV$, and at least two jets with $\abs{\eta}<4.7$ and $\pt>40\,(30)\GeV$ for the leading (second) jet.
A separation of $\DR>0.5$ is required between any two selected objects (photon, lepton, jets), as detailed in Section~\ref{sec:fidewxsmeas}.
In the electron channel we further require the invariant mass ($\mlepg$) of the selected photon and electron to be inconsistent with the \PZ~boson mass peak, $\abs{\mlepg-91}>10\GeV$, which suppresses the $\PZ\to\Pep\Pem$ background where one electron is misidentified as a photon.
Based on the pseudorapidity of the photon, the electron and muon channels are each subdivided into a barrel region with $\abs{\etag}<1.444$, and an endcap region with $1.566<\abs{\etag}<2.5$.

In this analysis, both a control and a signal region are defined.
The control region (CR) is constructed with an aim of validating the simulated samples and background estimation methods using data.
In addition to the previous selections, the control region is defined by a requirement that $200<\mjj<400\GeV$.

The signal region (SR) is defined by the previous selections plus the additional requirements that $\mjj>500\GeV$, $\abs{\dejj}>2.5$, $\mWg>100\GeV$, $\kinsysY<1.2$~\cite{zeppenfeld}, and $\kinsysPhi>2$ radians, where \mWg and \phiWg are, respectively, the invariant mass and azimuthal angle of the \PW~boson and \Pgg system, \phiDijet is the azimuthal angle of the dijet system, and $y_{j1(2)}$ is the rapidity of the leading (second) jet.
The longitudinal component of the neutrino momentum is estimated by solving the quadratic equation that constrains the mass of the charged lepton and neutrino system to the world-average value of the \PW~boson mass~\cite{PDG2018}.
As described in Ref.~\cite{Sirunyan2018}, when there are multiple solutions, the one with the smallest longitudinal momentum is chosen; if there are only complex solutions, the real part is chosen as the longitudinal momentum.
The requirements on \kinsysY and on \kinsysPhi are intended to ensure that the momentum of the \Wg system is balanced by that of the dijet system, which would be the case if there were no additional QCD radiation.
These selection requirements were chosen by optimizing the expected significance of the EW signal.

\section{Background estimation}
\label{sec:bkg_estimation}
\begin{figure}[h]
  \centering
    \includegraphics[width=0.49\textwidth]{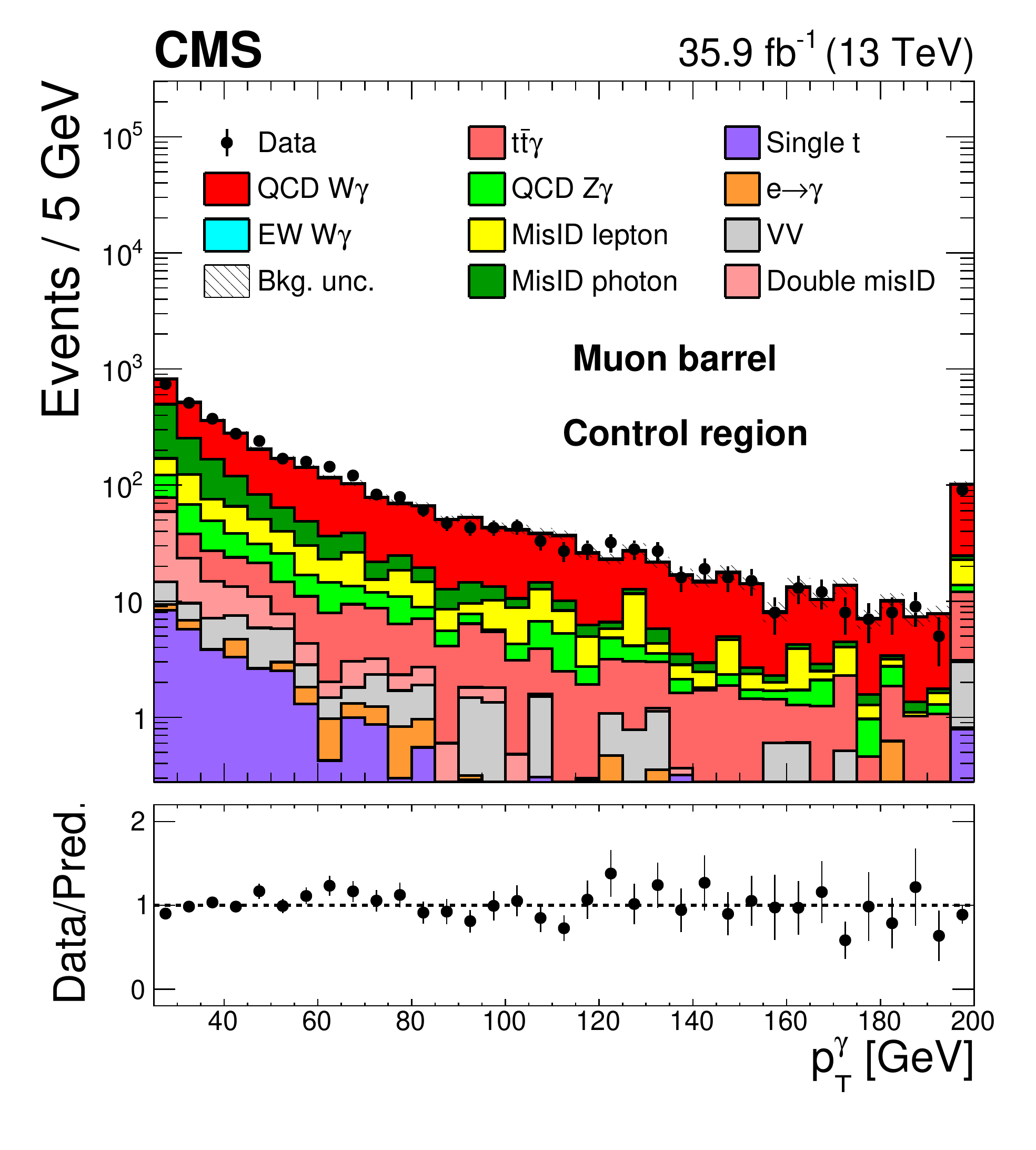}
    \caption{The photon \pt distribution in the muon barrel control region for data and background estimations.
      The misID backgrounds are derived from data, whereas the remaining backgrounds are estimated from simulation.
      All events with photon $\pt>195\GeV$ are included in the last bin.
      The hatched bands represent the statistical uncertainties on the predicted yields.
      The bottom graph shows the data divided by the prediction.}
    \label{fig:control}
\end{figure}

The backgrounds are shown in Fig.~\ref{fig:control}.
The yields of these backgrounds are obtained from a simultaneous fit to the data in both the SR and CR with the QCD \Wgjj normalization from the MC simulation. 
The theoretical and experimental uncertainties are assumed correlated between the SR and CR.  
The signal strength for the QCD \Wgjj background is $1.28^{+0.18}_{-0.16}$. 
The details are described in Sections~\ref{sec:systematics}--\ref{sec:fidewqcdxsmeas}.
Additional backgrounds are described in the following paragraphs.”  

Reconstructed photons and leptons that do not arise from outgoing particles in the hard interaction in the event are denoted as misidentified (misID) photons and leptons.
This category includes physical photons and leptons, as well as those of purely instrumental origins.
Because of the variety of sources of these misID particles and the difficulty of modeling instrumental effects, we use data-based methods to estimate their contribution.

The background from misID photons arises mainly from \Wpjets or top quark+jets events with a jet misreconstructed as a photon.
The method used to estimate this background involves measuring in CMS data and applying a per-photon extrapolation factor in which the denominator is chosen to be orthogonal to the full photon selection, but similar enough that the systematic uncertainties due to the extrapolation are well understood.
The photon in the denominator is required to fail the high-quality ID and pass the loose ID~\cite{Khachatryan:2017jub,wgzg}.
The extrapolation factor is determined from a template fit to the photon \sigmaee distribution, which is small for prompt photons and large for nonprompt photons.
The nonprompt template used in the fit is obtained from a sideband of the photon isolation variable in \Wpjets data.
More details can be found in Ref.~\cite{Sirunyan:2020tlu}.

The background from jets misidentified as leptons is estimated in a similar fashion.
To extrapolate from the loose leptons to the high-quality ones, an extrapolation factor is defined as:
\begin{linenomath}
\begin{equation*}
\frac{\frlep}{1 - \frlep},
\end{equation*}
\end{linenomath}
where \frlep is the lepton misidentification rate, defined as the ratio of the number of misID leptons where the lepton passes the high-quality ID to the total number passing only the loose ID requirements.
To reduce additional contamination from genuine leptons, the \Wpjets and \Zpjets contributions are subtracted from both the numerator and denominator.
The extrapolation factor is measured as a function of the $\eta$ and \pt of the lepton in a CR dominated by dijet events.
The dijet CR is defined by selecting one lepton, one jet that is well separated from the lepton, and low \ptmiss.
This technique is also used and described in Ref.~\cite{Sirunyan:2017ret}.

The background category ``double misID'' is defined as events containing both a misID photon and a misID lepton.
Its yield is estimated from a sample where both the photon and the lepton that are required to pass the loose ID selection, and fail the high-quality ID.
Such events are assigned a weight equal to the product of the misID extrapolation factors of the photon and lepton.
Double misID events contaminate the single misID background estimate because the second object is assumed to be genuine.
Consequently, each time a weight is added to the double misID estimate, the same weight is subtracted from both the single-photon and single-lepton estimates.
In addition, events in which genuine photons and leptons pass the loose ID but fail the high-quality ID selection contaminate both the single and double misID estimates. This source of contamination is estimated and removed using simulated events with reconstructed objects matched to generator-level objects.

Other backgrounds, including top~quark and diboson processes, are estimated from MC simulation and are normalized to the integrated luminosity of the CMS data set using inclusive cross sections calculated at NLO in QCD.
The \eletogam background includes events with an electron misID as a photon.
We apply $\abs{\mlepg-91} > 10\GeV$ to minimize this contribution.
The remaining background is estimated from simulated Drell--Yan and \ttg events that contain a photon matched to an electron at the generator level with $\DR=0.3$.

Fig.~\ref{fig:control} shows the photon pt distribution in the muon barrel control region for data and background estimates. The data and the estimates are in good agreement.

\section{Systematic uncertainties}
\label{sec:systematics}
Systematic uncertainties that affect the measurements arise from experimental inputs, such as detector effects and the methods used to compute higher-level quantities, \eg, efficiencies, and theoretical inputs such as the choice of the renormalization (\muR) and factorization (\muF) scales, and the choice of PDF sets.
Each source of systematic uncertainty is quantified by evaluating its effect on the yield and distribution of relevant kinematic variables in the signal and background categories.
The uncertainties are propagated to the final distributions and calculated bin-by-bin as described in Section~\ref{sec:ewsig}.

Table~\ref{tab:uncertainties} summarizes all the systematic uncertainties. The systematic uncertainties in the lepton trigger, reconstruction, and selection efficiencies, measured using a tag-and-probe technique, are 2--3\%.
The uncertainties in jet energy scale (JES) have the largest impact on the measurement.
The JES and jet energy resolution (JER) effects are estimated by shifting/smearing the jets in the simulations up and down by one standard deviation, and are then propagated to all relevant variables including VBS jet kinematic properties and \ptmiss, based on which the impact on signal and background yields are evaluated.
The uncertainties due to the JES and JER corresponding to different processes and different \mjj-\mlepg bins are in the ranges 0.9--78\% and 0.7--21\%, respectively.
An uncertainty of 2.5\% in the integrated luminosity~\cite{LUM-17-001} is used for all processes estimated from simulation and for the specified fiducial cross section.
The statistical uncertainties due to the finite size of both the simulated and data samples used in our background and signal prediction are estimated assuming Poisson statistics.
The uncertainties related to the finite number of simulated events or to the limited number of events in the data control samples are 7--11\% for the EW \Wgjj signal, 6--36\% for the QCD-induced \Wg background, 43--72\% for the nonprompt lepton contamination and 7--36\% for the nonprompt photon background.
These uncertainties are uncorrelated across different processes and bins of any single distribution, and grow with increasing \mjj and \mlepg.

An overall systematic uncertainty in the nonprompt photon background estimate is defined as the quadratic sum of the systematic uncertainties from several distinct sources.
An uncertainty because of the choice of the isolation variable sideband is evaluated by estimating the nonprompt photon fraction with alternative choices of the isolation sideband~\cite{Khachatryan:2017jub}.
A nonclosure uncertainty is defined by performing the nonprompt photon fraction fits using simulated events and comparing the results with the known fractions.
The nonclosure uncertainty in the endcap region is worse than in the barrel region and worsens as the photon \pt increases.
The overall systematic uncertainty in the nonprompt photon background is in the range of 12--22\%, dominated by the nonclosure.
Similarly, the dominant uncertainty in the nonprompt lepton estimate is associated with the nonclosure, which is calculated by comparing two yields, one from the \gpjets events and the other from the \gpjets events where the misID lepton rates are applied to events with a lepton that passes the loose, but fails the high-quality ID.
The selection used is the same as in the main event selection, except that the \mTW and \ptmiss requirements are removed to increase the statistical power.
The uncertainty associated with the nonprompt lepton background is 30\%.

The effects of the choice of \muR and \muF in the theoretical calculation for signal and background processes are estimated by independently changing \muR and \muF up and down by a factor of two from their nominal value in each event, with the condition that $1/2<\muR/\muF<2$.
The uncertainties are defined as the maximal differences from the nominal values.
The PDF uncertainties are evaluated according to the procedure described in Ref.~\cite{nnpdf_4lhc} using the NNPDF 3.0 set.
For the signal process, the scale uncertainty varies within the range of 1.5--11\% and the PDF uncertainty varies within the range 3.2--5.6\%, increasing with \mjj and \mlepg.
The scale uncertainty in the QCD-induced \Wg process, which has a very large impact on the measurement, varies in the range 6.1--20\%.
It is constrained by the simultaneous fit to the data in the CR.
The PDF uncertainty of QCD-induced \Wg production is in the range of 1--2\%.

The interference term between the EW- and QCD-induced processes, \ie, $\mathcal{O}(\alpha^4\alpS)$ at tree level, is estimated at particle level using MG5.
The contribution of the interference is calculated as the difference between the inclusive \Wgjj production, which contains the interference term, and the sum of the pure EW- and QCD-induced \Wgjj.
The interference is positive, and the ratio of the interference to EW \Wgjj is in the range 2--4\%, decreasing with increasing \mjj.
These values are used as systematic uncertainties in the signal process.

A correction factor is applied to the simulated events to account for the L1 trigger occasionally firing at the wrong time because of the darkening of the ECAL crystals.
This mistiming results in a loss of trigger efficiency in the data and is not modeled by the simulation.
The uncertainties due to these correction factors vary by 1--4\%, and are treated as correlated across different processes and bins.

\begin{table*}[htbp]
  \topcaption{Relative systematic uncertainties in the estimated signal and background yields in units of percent.
    The ranges reflect the dependence of the specified uncertainty on \mjj and \mlepg.}
\centering
\cmsTable{
\begin{tabular}{lcccccccccc}
\hline
Source	&  EW \Wgjj &  QCD \Wgjj &  \VV	&  \ttg	&  QCD \Zg &  Single \cPqt &  {\begin{tabular}[c]{c}MisID\\ photon\end{tabular}}  &  {\begin{tabular}[c]{c}MisID\\ lepton\end{tabular}} &  {\begin{tabular}[c]{c}Double\\ misID\end{tabular}} &  \eletogam \\
\hline
  JES				& 0.9--6.9  & 11--28    & 6.4--38  & 3.7--16  & 12--78    & 3.3--18  &   \NA    &    \NA    &   \NA   &  11--28     \\
  JER 				& 0.7--2.2  & 0.7--4.1  & 6.9--21  & 1.3--4.9 & 6.5--15   & 2.9--7.1 &   \NA    &    \NA    &   \NA   &  0.7--4.1   \\
  Integrated luminosity		&  2.5      &  2.5      &   2.5    &   2.5    &   2.5     &   2.5    &   \NA    &    \NA    &   \NA   &  2.5        \\
  MisID photon			&   \NA     &   \NA     &   \NA    &   \NA    &  \NA      &   \NA    & 12--22   &    \NA    &  12--22 &   \NA       \\
  MisID lepton			&   \NA     &   \NA     &   \NA    &   \NA    &  \NA      &   \NA    &   \NA    &   30      &  30     &   \NA       \\
  \muR/\muF scales 		&  1.5--11  & 6.1--20   &   \NA    &   \NA    &  \NA      &   \NA    &   \NA    &    \NA    &   \NA   &   \NA       \\
  PDF				&  3.2--5.6 &   1--2    &   \NA    &   \NA    &  \NA      &   \NA    &   \NA    &    \NA    &   \NA   &   \NA       \\
  Interference			&  1.8--2.8 &   \NA     &   \NA    &   \NA    &  \NA      &   \NA    &   \NA    &    \NA    &   \NA   &   \NA       \\
  Cross section for \ttg	&   \NA     &   \NA     &   \NA    &    10    &  \NA      &   \NA    &   \NA    &    \NA    &   \NA   &   \NA       \\
  Cross section for \VV		&   \NA     &   \NA     &    10    &   \NA    &  \NA      &   \NA    &   \NA    &    \NA    &   \NA   &   \NA       \\
  Modeling of pileup		&  0--0.6   &  0.3--1.4 & 4.8--13  & 2.6--3.9 & 6.2--19   & 1.0--3.9 &   \NA    &    \NA    &   \NA   &  0.3--1.4   \\
  Statistical uncertainty	&  7--11    &  6--36    & 45--100  & 13--56   & 16--100   & 17--55   & 7--36    & 43--72    & 30--100 &  54--100    \\
  L1 mistiming                  &  1.7--2.4 &  0.8--1.6 & 0.5--1.6 & 1.4--2.5 & 0.6--3.6  & 1.0--2.1 &   \NA    &    \NA    &   \NA   &  1.1--2.8   \\
  Muon ID/Iso			&  0.3      &  0.3      &  0.3     &   0.3    &   0.3     &   0.3    &   \NA    &    \NA    &   \NA   &   0.3       \\
  Muon trigger			&  0.3      &  0.2      &  0.2     &   0.2    &   0.1     &   0.1    &   \NA    &    \NA    &   \NA   &   0.2       \\
  Electron reconstruction	&  0.5      &  0.6      &  0.5     &   0.6    &   0.6     &   0.5    &   \NA    &    \NA    &   \NA   &   0.5       \\
  Electron ID/Iso		&  1.3      &  1.3      &  1.3     &   1.3    &   1.3     &   1.3    &   \NA    &    \NA    &   \NA   &   1.3       \\
  Electron trigger		&  2.5      &  2.5      &  2.5     &   2.5    &   2.5     &   2.5    &   \NA    &    \NA    &   \NA   &   2.5       \\
  Photon ID			&  1.2      &  1.2      &  1.1     &   1.2    &   1.3     &   1.2    &   \NA    &    \NA    &   \NA   &   1.2       \\
\hline
\end{tabular}
}
\label{tab:uncertainties}
\end{table*}

All of the systematic uncertainties discussed above are applied both to the signal significance measurement and in the search for aQGC contributions.
They are also propagated to the uncertainty in the measured fiducial cross section, with the exception of the theoretical uncertainties associated with the signal cross section.
All of the systematic uncertainties except those that arise from the trigger efficiency and the lepton identification and misidentification are considered to be correlated between the electron and muon channels.

\section{The EW \texorpdfstring{\Wg}{W gamma} production measurement}
\label{sec:ewsig}

\begin{figure*}[htbp]
  \centering
      \includegraphics[width=0.49\textwidth]{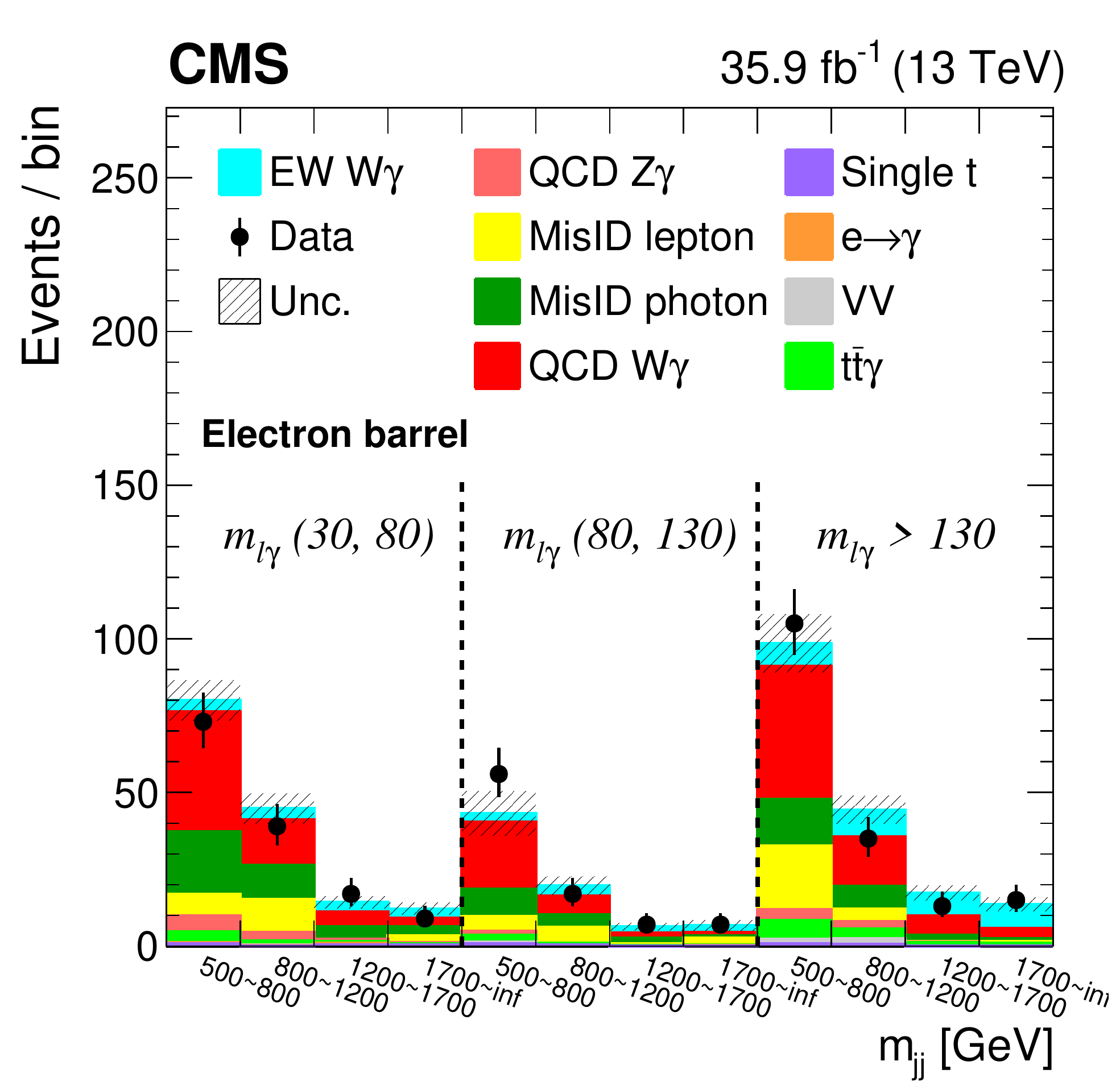}
      \includegraphics[width=0.49\textwidth]{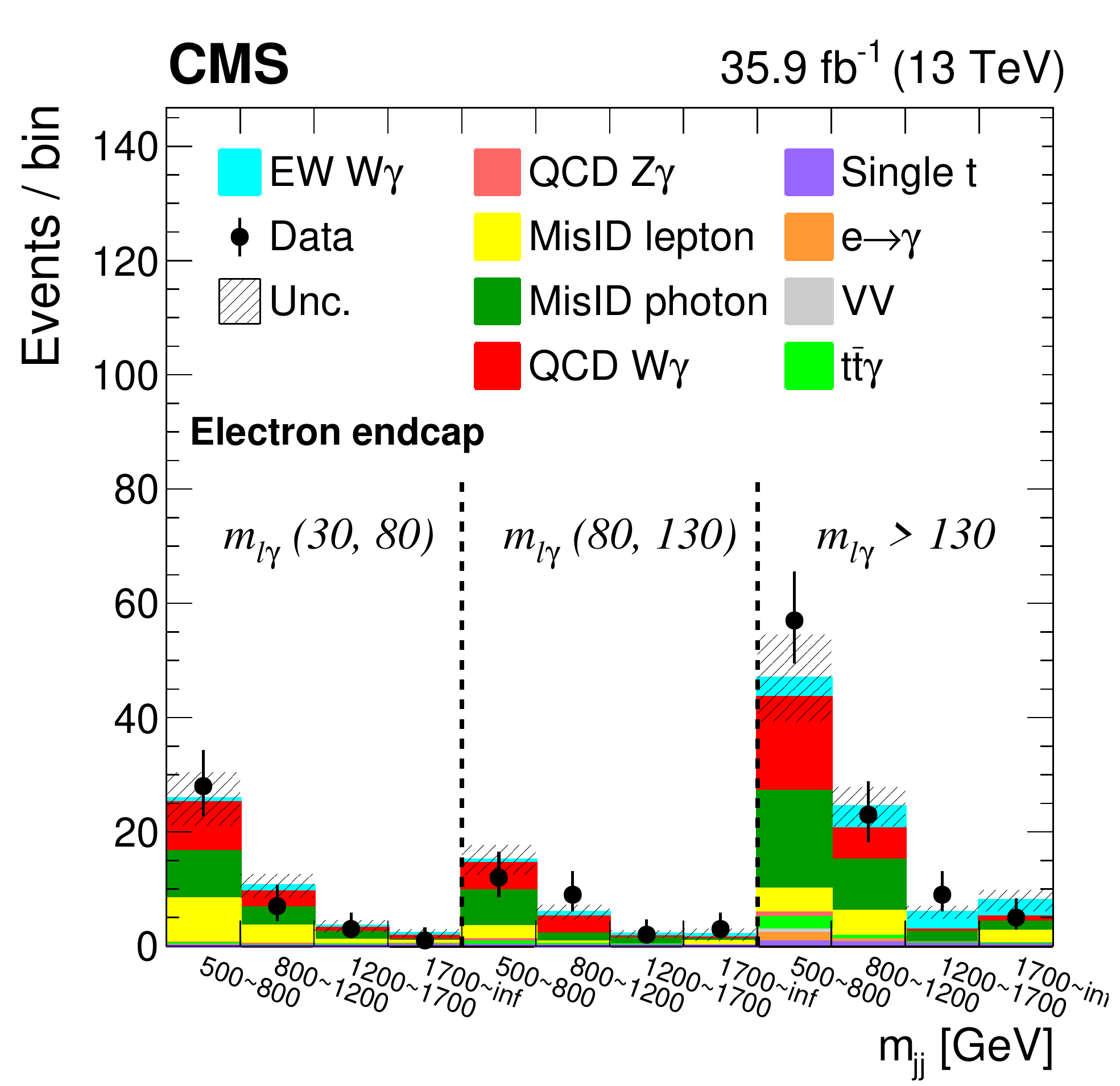}
      \\
      \includegraphics[width=0.49\textwidth]{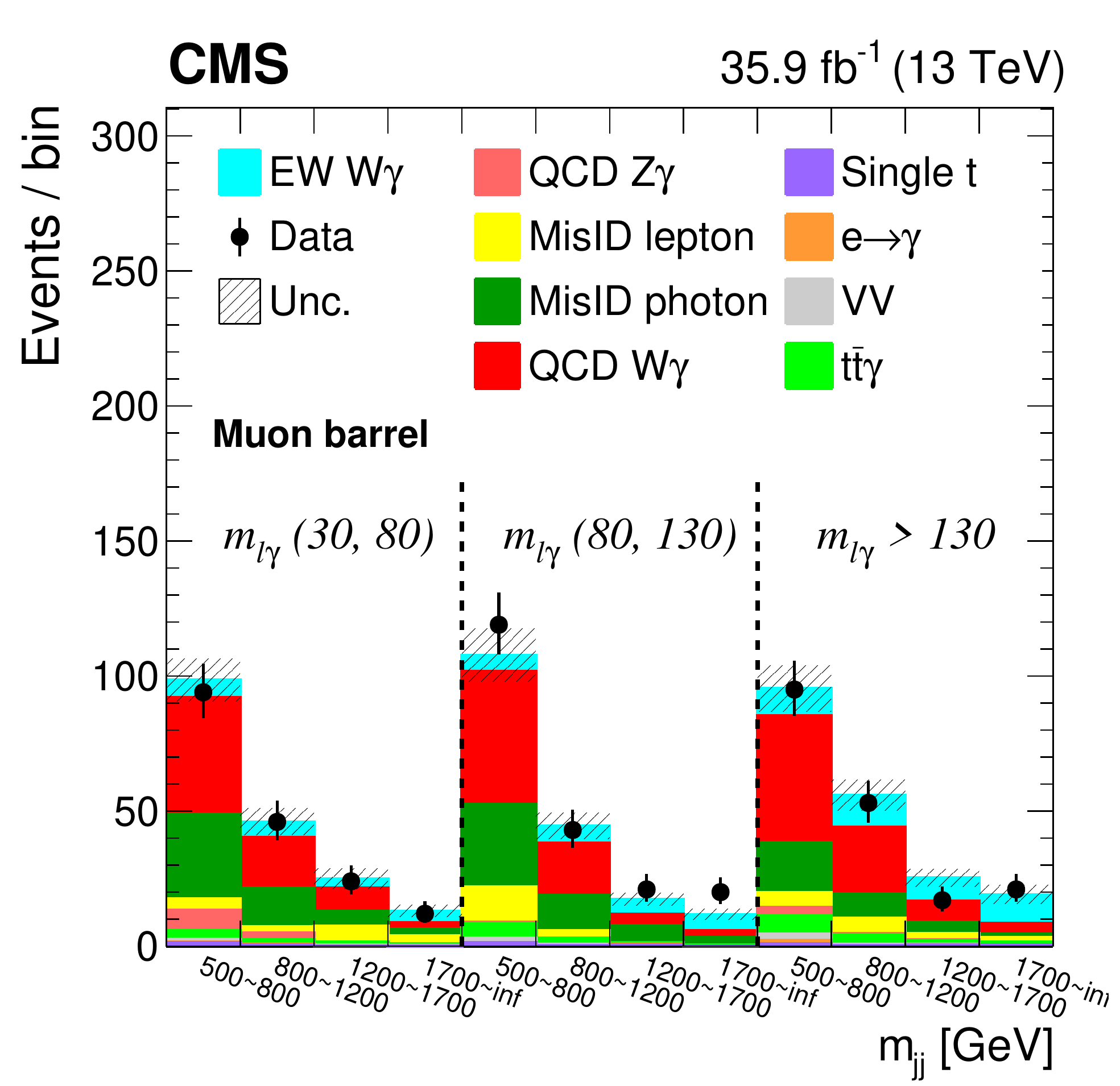}
      \includegraphics[width=0.49\textwidth]{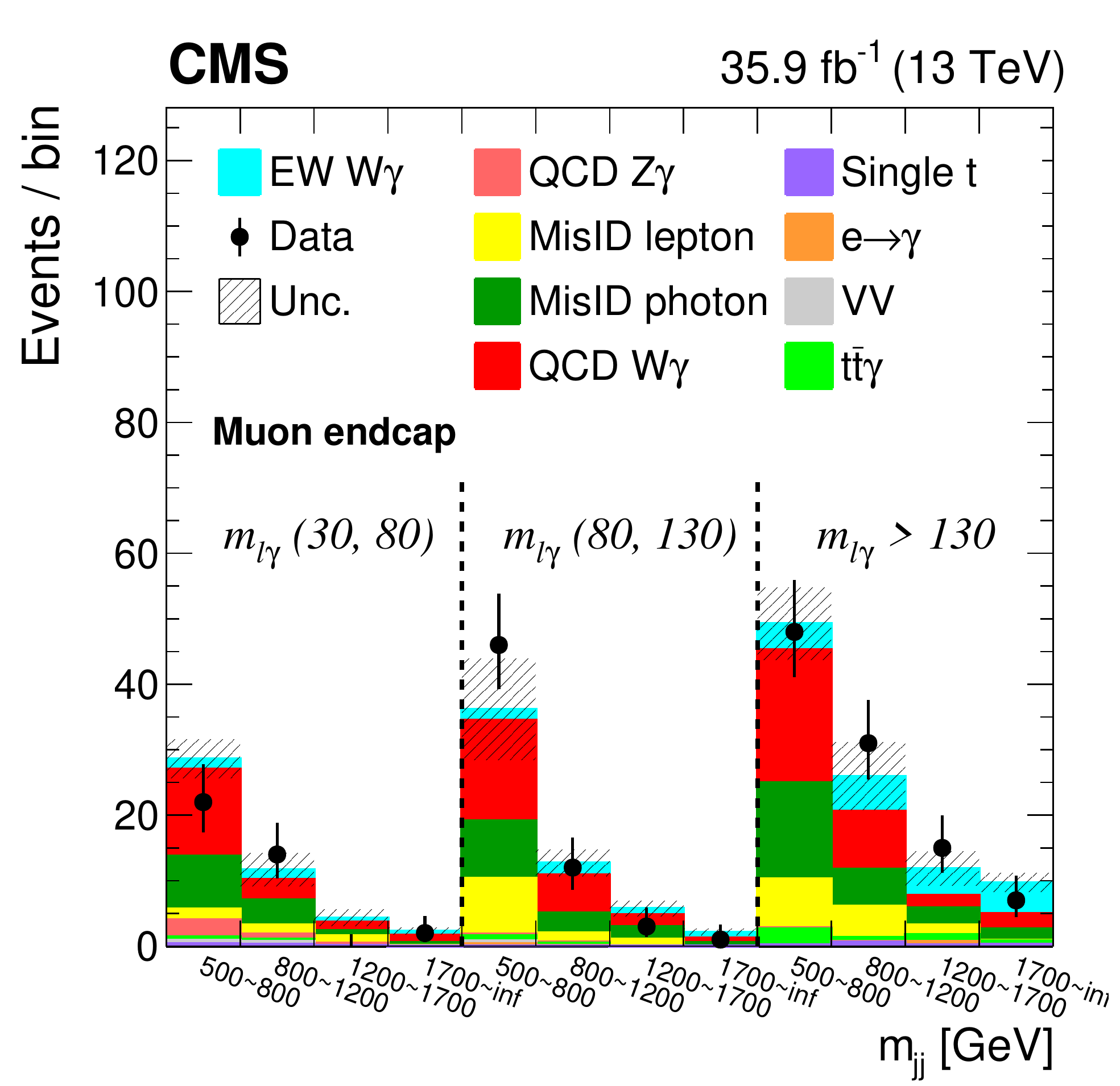}
      \caption{The 2D distributions used in the fit for the signal strength of EW {\Wg}+2 jets for events in the electron barrel (upper left), electron endcap (upper right), muon barrel (lower left), and muon endcap (lower right).
		The hatched bands represent the systematic uncertainties on the predicted yields.
		The predicted yields are shown with their best-fit normalizations.
		}
      \label{fig:2d_strength}
\end{figure*}

Table~\ref{tab:yields} shows the simulated signal and background yields prior to any fitting, as well as the observed data yields.
To quantify the significance of the observation of EW production of the \Wg signal, we perform a statistical analysis of the event yields through a fit to the (\mjj,\mlepg) two-dimensional (2D) distribution.
Both \mjj and \mlepg are powerful variables for distinguishing between the signal and QCD \Wg background, and the 2D analysis provides a larger expected significance than either variable alone.
For this measurement, and the measurements in Sections~\ref{sec:fidewxsmeas} and~\ref{sec:fidewqcdxsmeas}, the SR is further divided into four bins in \mjj (lower boundaries of 500, 800, 1200, and 1700\GeV) and three bins in \mlepg (lower boundaries of 30, 80, and 130\GeV).
The data in the CR are fit simultaneously with the data in the SR.
Figure~\ref{fig:2d_strength} shows the resultant 2D fitted distributions.

\begin{table*}[htbp]
\centering
  \topcaption{Signal, background, and data yields after the final selection.
    Statistical and systematic uncertainties (before the fitting) are added in quadrature.}
\begin{tabular}{lcccc}
\hline
                        &  Electron barrel   &  Electron endcap &  Muon barrel        &  Muon endcap       \\
\hline
MisID photon 		& $ 81.0 \pm  5.2 $  & $ 48.1 \pm 4.9 $ & $ 134.8 \pm  8.2 $  & $ 52.1 \pm  4.8 $  \\
MisID lepton 		& $ 63.7 \pm 12.3 $  & $ 27.8 \pm 7.2 $ & $  46.8 \pm 10.6 $  & $ 23.1 \pm  6.5 $  \\
QCD \Wgjj               & $154.2 \pm 12.0 $  & $ 41.1 \pm 4.4 $ & $ 221.2 \pm 15.8 $  & $ 72.1 \pm  6.2 $  \\
\ttg			& $ 20.6 \pm  1.6 $  & $  5.1 \pm 0.6 $ & $  28.3 \pm  1.8 $  & $  6.9 \pm  0.8 $  \\
QCD \Zg			& $ 18.0 \pm  3.1 $  & $  1.9 \pm 0.9 $ & $  16.2 \pm  3.0 $  & $  4.9 \pm  1.3 $  \\
Single \cPqt		& $  4.9 \pm  0.8 $  & $  2.5 \pm 0.5 $ & $   6.8 \pm  0.9 $  & $  2.4 \pm  0.5 $  \\
\VV			& $  4.2 \pm  1.6 $  & $  0.6 \pm 0.6 $ & $   7.5 \pm  2.1 $  & $  1.4 \pm  0.7 $  \\
\eletogam               & $  1.5 \pm  0.6 $  & $  2.1 \pm 0.8 $ & $   1.7 \pm  0.7 $  & $  1.1 \pm  0.6 $  \\
[\cmsTabSkip]
Total background        & $348.3 \pm 18.4 $  & $129.1 \pm 9.9 $ & $ 463.4 \pm 21.2 $  & $163.8 \pm 10.4 $  \\
[\cmsTabSkip]
EW \Wgjj                & $ 48.8 \pm  2.2 $  & $ 16.1 \pm 1.0 $ & $  74.5 \pm  2.8 $  & $ 24.4 \pm  1.3 $  \\
[\cmsTabSkip]
Total predicted		& $397.1 \pm 18.5 $  & $145.2 \pm 10.0$ & $ 537.9 \pm 21.4 $  & $188.2 \pm 10.5 $  \\
[\cmsTabSkip]		
Data                    & 393                & 159              &     565             &   201              \\
\hline
\end{tabular}
\label{tab:yields}
\end{table*}

The signal significance is quantified on the basis of a profile likelihood test statistic~\cite{Wilks:1938dza}.
This test statistic involves the ratio of two Poisson likelihood functions, one in which the signal strength is fixed to zero and one in which the signal strength is allowed to have any positive value.
The signal strength represents the ratio of observed to expected signal yields.
Systematic uncertainties are added as parameters into the likelihood function to scale the relevant process using log-normal functions.
The distribution in the test statistic is assumed to be in the asymptotic regime where there is a simple relationship between its value and the significance of the result~\cite{CLs}.
The observed (expected) signal strength parameter is $\muhat = 1.20^{+0.26}_{-0.24}\,(1.00^{+0.27}_{-0.25}$), corresponding to an observed (expected) statistical significance of 4.9 (4.6)\sdev for the analyzed 13\TeV data set.

This result can be combined with the previous CMS measurement at 8\TeV described in Ref.~\cite{Khachatryan:2016vif} assuming the signal strength does not change with the center of mass energy.
There are two uncertainties that are correlated between the 8 and 13\TeV analyses.
The theoretical uncertainties in the signal and QCD \Wg background of the 8\TeV analysis include multiple sources, but are dominated by the renormalization and factorization scale uncertainties, and are therefore correlated with the corresponding uncertainties in the 13\TeV analysis.
All other uncertainties are uncorrelated between the 8 and 13\TeV analyses. After combining our result with that at 8\TeV using this correlation scheme, the observed (expected) significance is 5.3 (4.8)\sdev.

\section{Fiducial EW \texorpdfstring{\Wgjj}{W gamma jj} cross section measurement}
\label{sec:fidewxsmeas}
A fiducial cross section at 13\TeV is extracted in the same (\mjj,\mlepg) binning used in the calculation of significance, and through the same simultaneous fit used in the CR.
The fiducial region is defined using the MC generator quantities: one lepton with $\ptlep>30\GeV$ and $\abs{\etalep}<2.4$, $\ptmiss>30\GeV$,  $\ptg>25\GeV$, $\abs{\etag}<1.444$ or $1.566<\abs{\etag}<2.5$, $\DR_{\Pell\gamma}>0.5$, $\mTW>30\GeV$, two jets with $\ptX{j1(2)}>40\,(30)\GeV$, with $\abs{\etaj}<4.7$, $\mjj>500\GeV$, $\DR_{\jj}>0.5$, $\DR_{\text{j}\Pell}>0.5$, $\DR_{\text{j}\gamma}>0.5$, and $\abs{\dejj} > 2.5$.
The leptons are reconstructed at the particle level with fully recovered final-state radiation.
The acceptance is defined as the fraction of the generated signal events passing the fiducial region selection, which is extracted using MG5.
The theoretical uncertainty because of the extrapolation between the fiducial and SR is negligible ($<1\%$).
We define the cross section as
\begin{linenomath}
\begin{equation*}\label{eq:sigfid}
\sigfid (13\TeV) = \sigmag\,\muhat\,\alphagf,
\end{equation*}
\end{linenomath}
where the cross section for the generated signal events is $\sigmag = 0.776\unit{pb}$, the signal strength parameter $\muhat = 1.20^{+0.26}_{-0.24}$, and the acceptance $\alphagf = 0.02195$.
The observed fiducial cross section is $\sigfidew(13\TeV) = 20.4\pm 0.4\lum\pm 2.8\stat\pm 3.5\syst\unit{fb} = 20.4\pm 4.5\unit{fb}$.

\section{Fiducial EW+QCD \texorpdfstring{\Wgjj}{W gamma jj} cross section measurement}
\label{sec:fidewqcdxsmeas}
In addition to the EW {\Wgjj} process, we also determine a cross section for inclusive EW+QCD {\Wgjj} production.
The fiducial region is the same as that for EW {\Wgjj} and the formula for the cross section is
\begin{linenomath}
\begin{equation}
  \label{eq:sigfideq}
\sigfid = \mu \left\{\sigmagew \alphaewgf+\sigmagqcd \alphaqcdgf\right\}. \nonumber
\end{equation}
\end{linenomath}
Since the QCD {\Wg}+2 jets is part of the signal, the CR is no longer included in the calculated signal strength.

\begin{figure*}[htbp]
  \centering
      \includegraphics[width=0.49\textwidth]{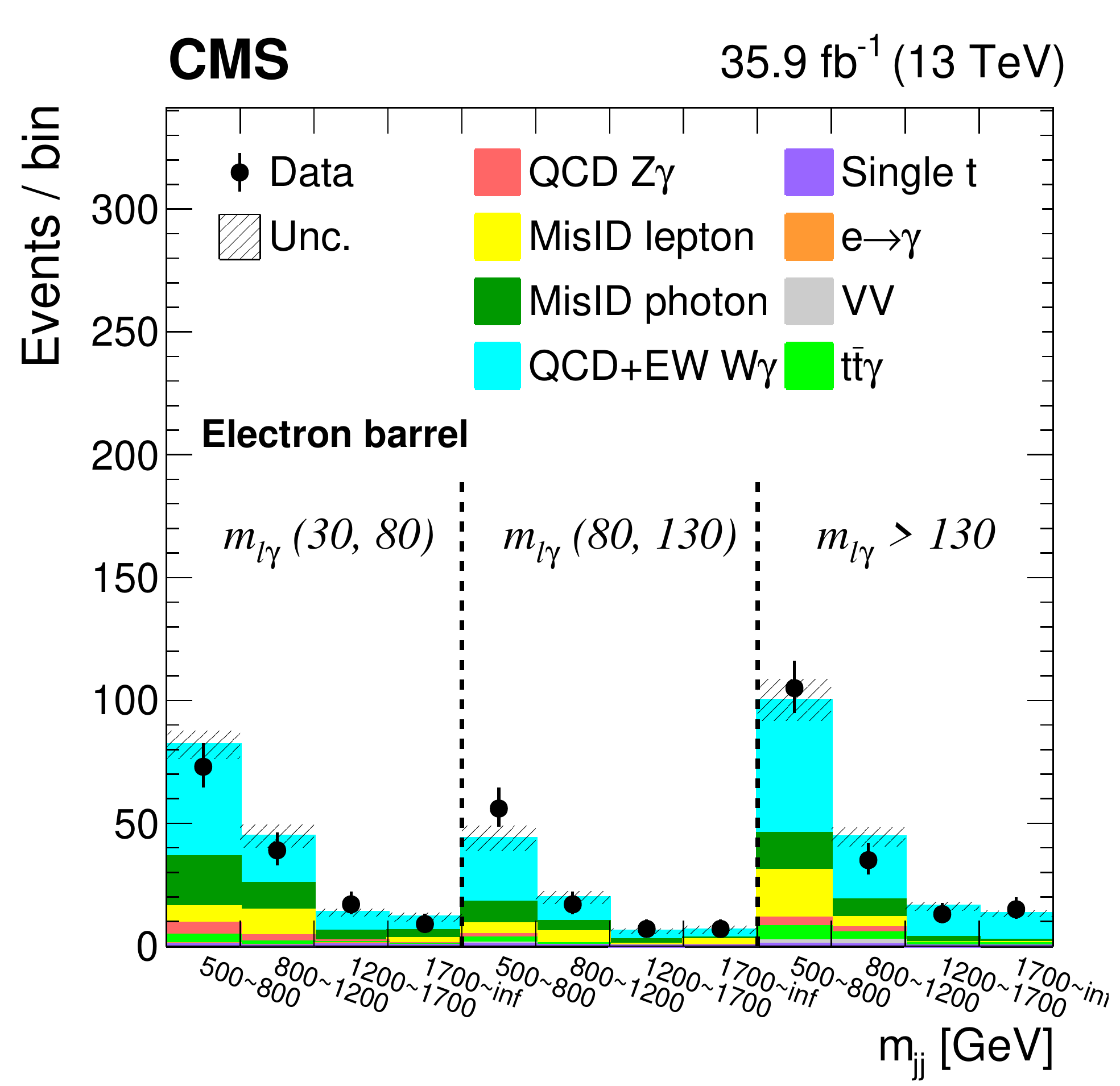}
      \includegraphics[width=0.49\textwidth]{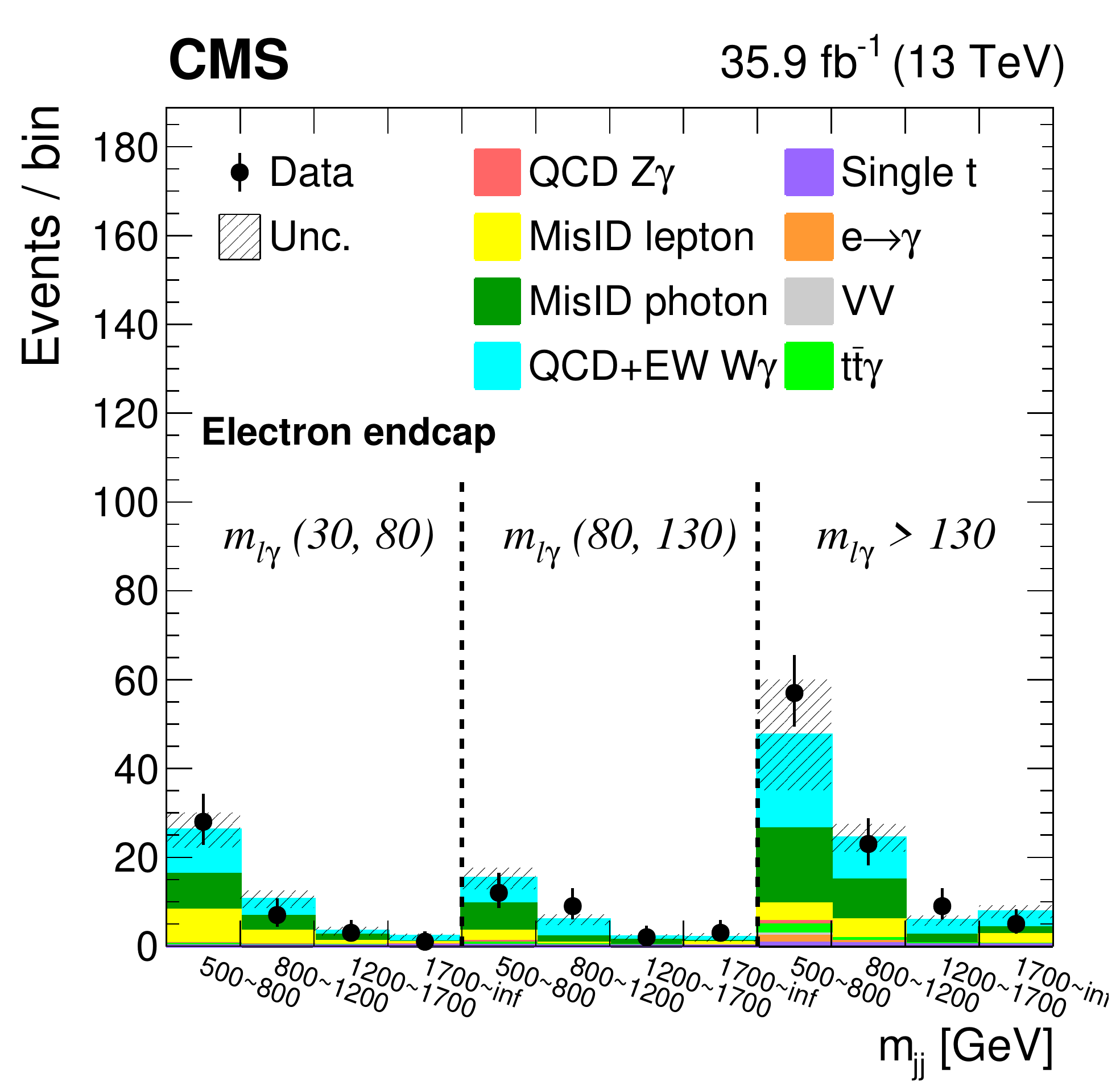}
      \\
      \includegraphics[width=0.49\textwidth]{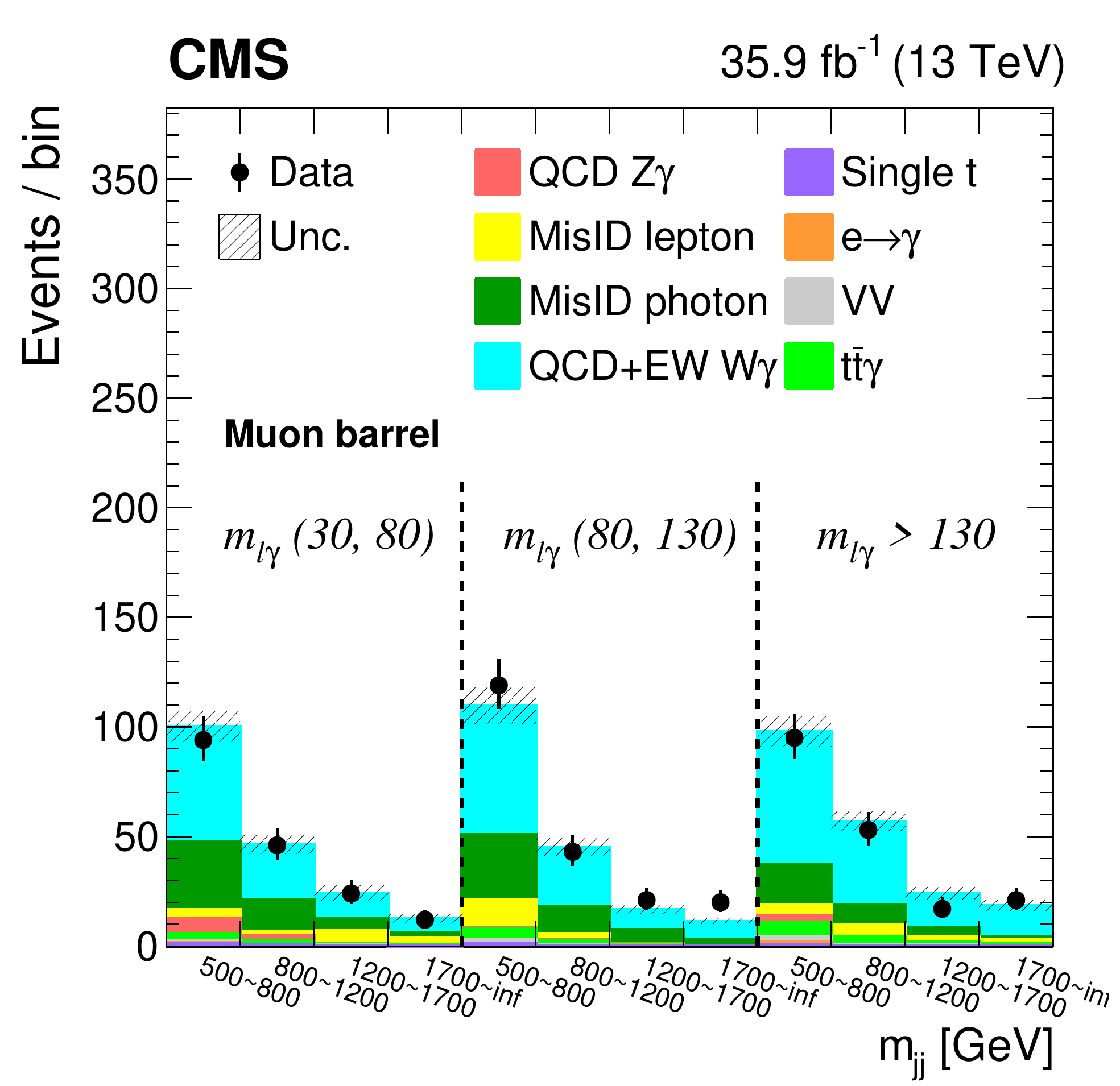}
      \includegraphics[width=0.49\textwidth]{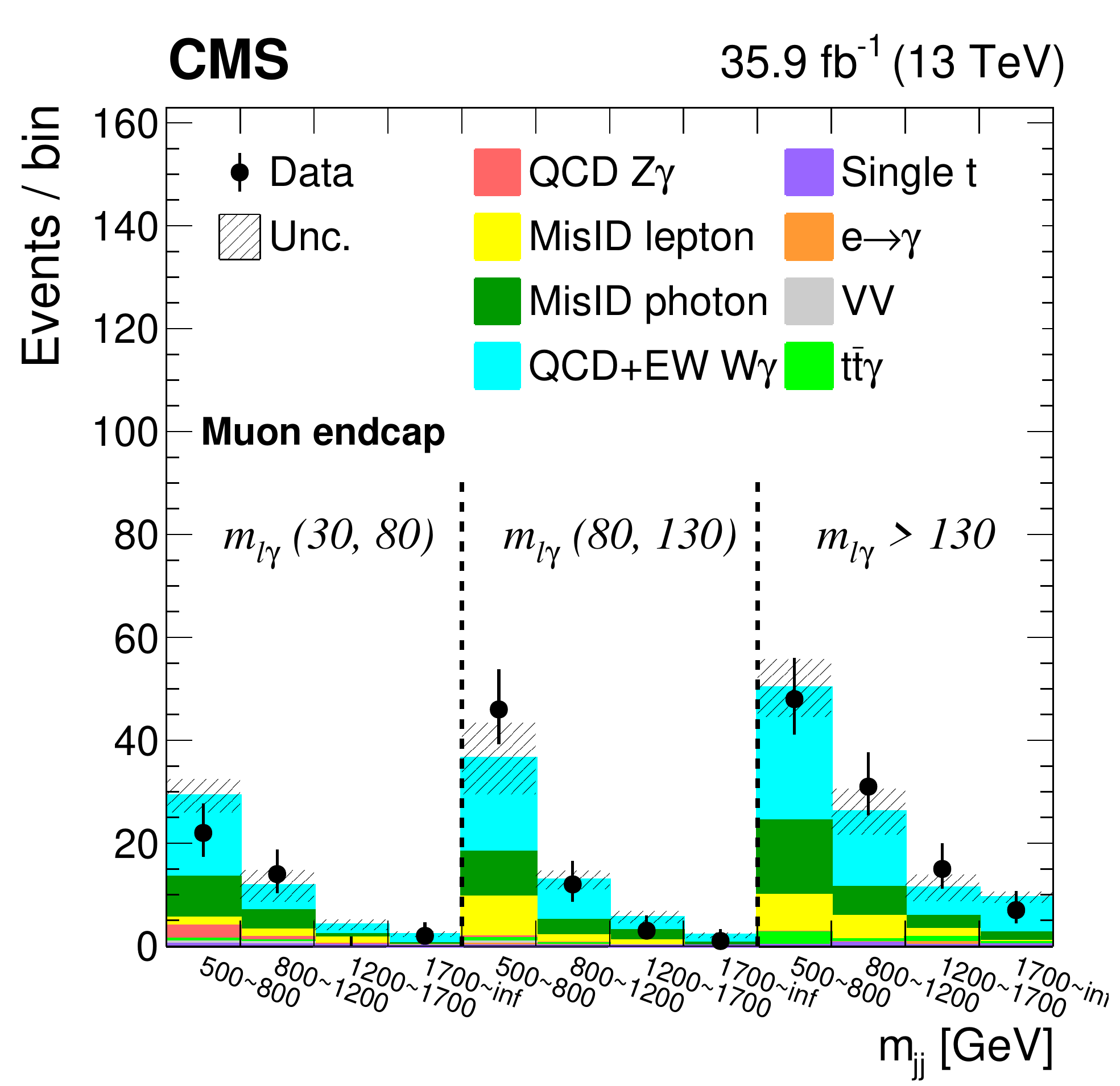}
      \caption{The 2D distributions used in the fit for the signal strength of EW+QCD {\Wg}+2 jets in the electron barrel (upper left), electron endcap (upper right), muon barrel (lower left) and muon endcap (lower right).
		The hatched bands represent the systematic uncertainties on the predicted yields.
		The predicted yields are shown with their best-fit normalizations.
		}
      \label{fig:2d_strength_inclusive}
\end{figure*}

The inputs used for the fit are similar to the ones for EW {\Wgjj}, with the difference that EW and QCD {\Wgjj} are combined as signal.
The cross section for QCD {\Wgjj} is 178.6\unit{pb}, and \alphaqcdgf is calculated to be 0.0004068.
The measured signal strength for inclusive {\Wgjj} is $1.21^{+0.17}_{-0.16}$ and the observed fiducial cross section is $\sigfidewq(13\TeV) = 108\pm 2\lum\pm 5\stat\pm 15\syst\unit{fb} = 108\pm 16\unit{fb}$.
Figure~\ref{fig:2d_strength_inclusive} shows the post-fit results.

\section{Limits on anomalous quartic gauge couplings}
\label{sec:aqgclimits}

The effects of BSM physics can be modeled in a generic way through a collection of linearly independent higher-dimensional operators in effective field theory~\cite{paper_aqgc}.
As mentioned above, VBS is more suitable to constrain aQGC. 
The lowest dimension operators that modify quartic gauge couplings but do not exhibit two or three weak gauge boson vertices are dimension-eight. 
Reference~\cite{Eboli:2006wa} proposes nine independent charge-conjugate and parity-conserving \\ 
dimension-eight effective operators by assuming the SU(2)$\times$U(1) symmetry of the EW gauge field.
The model includes a Higgs doublet to incorporate the presence of an SM Higgs boson.
A contribution from aQGCs enhances the production of events with large {\Wg} mass.
The operators affecting the {\Wgjj} channel can be divided into two categories.
The operators \loperator{M}{0}--\loperator{M}{7} contain an SU(2) field strength, the U(1) field strength, and the covariant derivative of the Higgs doublet field.
The operators \loperator{T}{0}--\loperator{T}{2} and \loperator{T}{5}--\loperator{T}{7}, contain only the two field strengths.
The coefficient of the operator \loperator{X}{Y} is denoted by \fcoefflam{X}{Y}, where $\Lambda$ is the unknown scale of BSM physics.

\begin{figure}[h]
  \centering
      \includegraphics[width=0.49\textwidth]{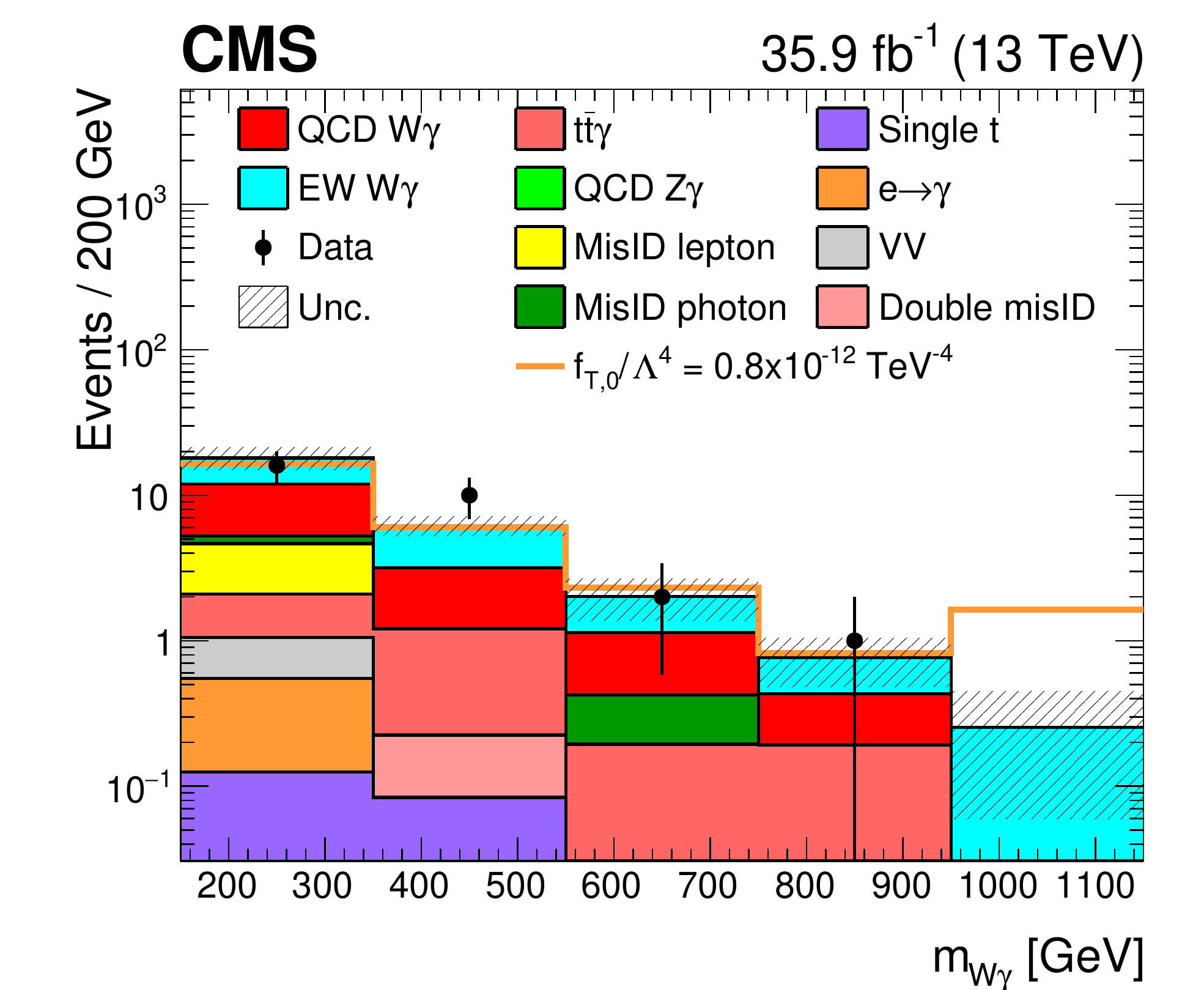}
      \caption{The \mWg distribution of events satisfying the aQGC region selection, which is used to set constraints on the anomalous coupling parameters.
        The orange line represents a nonzero \fcoefflam{T}{0} setting.
        All events with $\mWg > 950\GeV$ are included in the last bin.
	The hatched bands represent the statistical uncertainties in the predicted yields.}
      \label{fig:aqgc_mu}
\end{figure}

A simulation is performed that includes the effects of the aQGCs in addition to the SM EW {\Wgjj} process, as well as any interference between the two.
We use the \mWg distribution to extract limits on the aQGC parameters.
To obtain a continuous prediction for the signal as a function of the anomalous coupling, a quadratic fit is performed to the SM+aQGC yield as a function of the aQGC coefficient, separately in each \mWg bin in the aQGC region, which is defined based on the common selection in Section~\ref{sec:selection}, with the further requirements $\mjj > 800\GeV$, $\abs{\dejj} > 2.5$, $\mWg > 150\GeV$, and $\ptg > 100\GeV$.
As the aQGC contributions arise from pure VBS diagrams and are more enhanced in the VBS phase space region, and the anomalous operators lead to more energetic final state particles, the additional requirements are optimized to enhance the aQGC sensitivity based on the simulation studies.
Figure~\ref{fig:aqgc_mu} shows the resulting distribution in \mWg.
No statistically significant excess of events relative to the SM prediction is observed.

\begin{figure}[h!]
  \centering
      \includegraphics[width=0.49\textwidth]{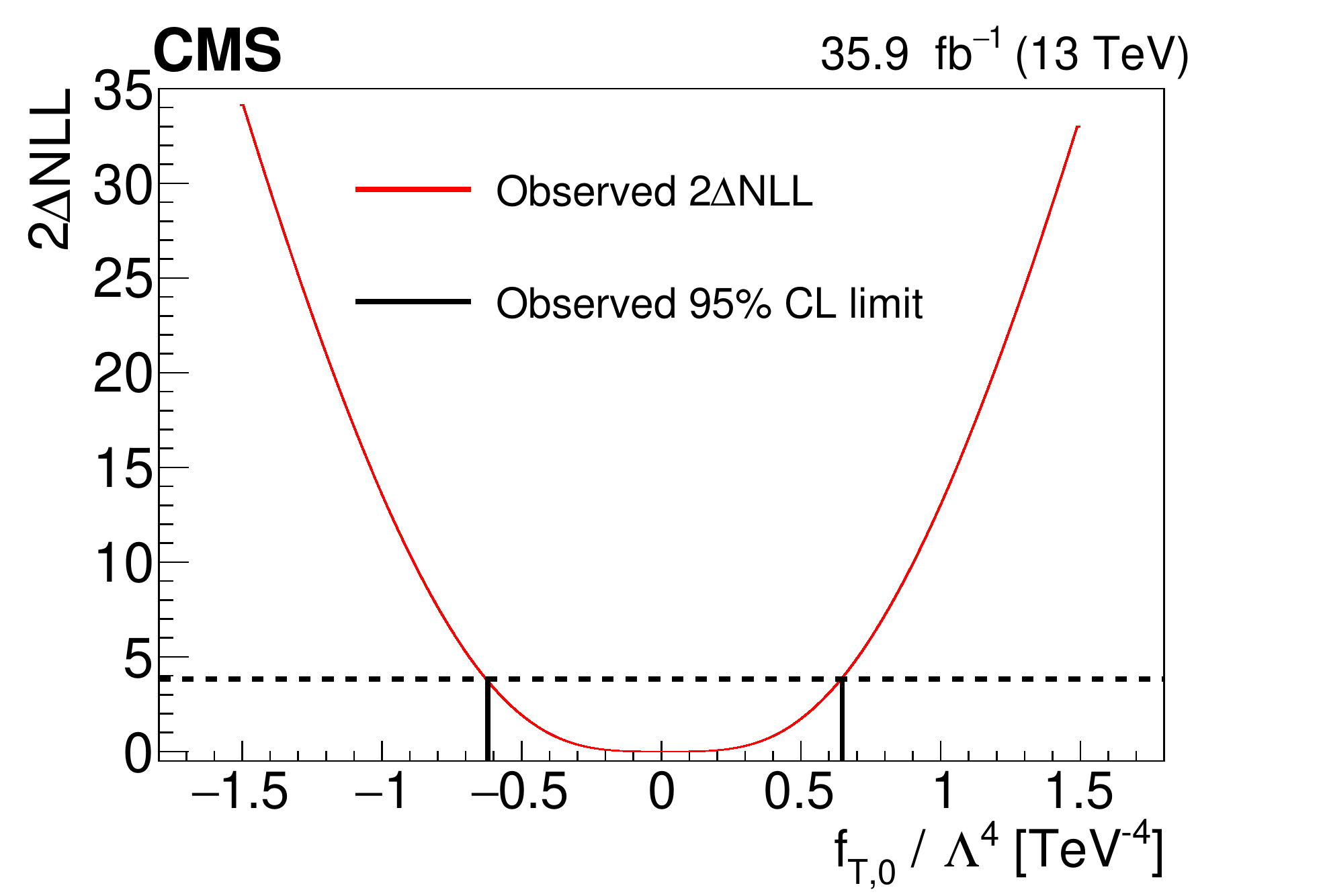}
      \caption{Observed 95\% \CL interval on the aQGC parameter \fcoefflam{T}{0}.}
      \label{fig:profile_expected}
\end{figure}

The following profile likelihood test statistic is used in the aQGC limit setting procedure:
\begin{linenomath}
\begin{equation*}\label{likelihoodformula}
t_{\alphatest} = -2 \log \frac{\mathcal{L}(\alphatest,{\thetabdhat})}{\mathcal{L}(\alphahat,\thetabhat)}.
\end{equation*}
\end{linenomath}
The likelihood function is the product of Poisson distributions and a normal constraining term with nuisance parameters representing the sources of systematic uncertainties in each bin.
The final likelihood function is the product of the likelihood functions of the electron and muon channels.
The main constraint on the aQGC parameters is from the highest \mWg bin.
The parameter \alphatest represents the aQGC point being tested, and the symbol \thetab represents a vector of nuisance parameters assumed to follow log-normal distributions.
The parameter \thetabdhat corresponds to the maximum of the likelihood function at the point \alphatest.
The \alphahat and \thetabhat parameters correspond to the single global maximum of the likelihood function.
This test statistic is assumed to follow a noncentral $\chi^2$ distribution~\cite{Wilks:1938dza}.
It is therefore possible to extract the limits immediately from the difference in the negative log-likelihood (NLL) function $\dnll = t_{\alphatest}/2$~\cite{Khachatryan:2014jba}.
The 95\% confidence level (\CL) limit on a one-dimensional aQGC parameter corresponds to $2\dnll = 3.84$.
Figure~\ref{fig:profile_expected} shows the likelihood scan of parameter \fcoefflam{T}{0} in the calculation of the observed limits.

The observed and expected 95\% \CL limits on the coefficients of these operators, shown in Table~\ref{tab:VBS_aQGC}, are obtained by varying the coefficient of one operator at a time, with all others set to 0, the SM value.
The yield of the EW signal in any bin is a quadratic function of the coefficient, whose minimum in general does not occur at a coefficient value of 0 because of interference with the SM operators.
We therefore set upper and lower limits on the operator coefficients through a limit-setting procedure that involves first obtaining the global maximum of the profile likelihood function, and then the maximum of the profile likelihood function at fixed coefficient values, which can be compared to the global maximum and converted to {\CL}s.
NLO EW corrections to VBS \Wg can be sizeable and increase as a function of \mjj, which may bias the aQGC measurement. Although there is no NLO EW calculation available yet for VBS \Wg,  we have instead tested with the numbers from same-sign WW scattering~\cite{Biedermann:2016yds,Biedermann:2017bss}, and the effect on the aQGC limit is found to be negligible.
The unitarity bound (\ubound) is defined as the scattering energy at which the aQGC coupling strength, when set equal to the observed limit, would result in a scattering amplitude that violates unitarity.
The value of \ubound is determined using the \textsc{vbfnlo}~2.7.1 framework~\cite{VBFNLO}, taking into account the difference between \textsc{vbfnlo} and MG5.
These are the most stringent limits to date on the aQGC parameters \fcoefflam{M}{2--5} and \fcoefflam{T}{6--7}.

\begin{table}[htb!]
\centering
\topcaption{The exclusion limits at 95\% \CL on each aQGC coefficient, parameterized using the distribution in \mWg, and listed along with the unitarity bound.
    All coupling parameter limits are in ${\TeVns}^{-4}$, while the \ubound values are in TeV.}
\begin{tabular}{cccc}
\hline
  Parameters & Obs. limit & Exp. limit & \ubound \\
\hline
  \fcoefflam{M}{0}  &  $[-8.1, 8.0]$  &  $[-7.7, 7.6]$  & 1.0  \\
  \fcoefflam{M}{1}  &  $[-12, 12]  $  &  $[-11, 11]$    & 1.2  \\
  \fcoefflam{M}{2}  &  $[-2.8, 2.8]$  &  $[-2.7, 2.7]$  & 1.3  \\
  \fcoefflam{M}{3}  &  $[-4.4, 4.4]$  &  $[-4.0, 4.1]$  & 1.5  \\
  \fcoefflam{M}{4}  &  $[-5.0, 5.0]$  &  $[-4.7, 4.7]$  & 1.5  \\
  \fcoefflam{M}{5}  &  $[-8.3, 8.3]$  &  $[-7.9, 7.7]$  & 1.8  \\
  \fcoefflam{M}{6}  &  $[-16, 16]$    &  $[-15, 15]$    & 1.0  \\
  \fcoefflam{M}{7}  &  $[-21, 20]$    &  $[-19, 19]$    & 1.3  \\
  \fcoefflam{T}{0}  &  $[-0.6, 0.6]$  &  $[-0.6, 0.6]$  & 1.4  \\
  \fcoefflam{T}{1}  &  $[-0.4, 0.4]$  &  $[-0.3, 0.4]$  & 1.5  \\
  \fcoefflam{T}{2}  &  $[-1.0, 1.2]$  &  $[-1.0, 1.2]$  & 1.5  \\
  \fcoefflam{T}{5}  &  $[-0.5, 0.5]$  &  $[-0.4, 0.4]$  & 1.8  \\
  \fcoefflam{T}{6}  &  $[-0.4, 0.4]$  &  $[-0.3, 0.4]$  & 1.7  \\
  \fcoefflam{T}{7}  &  $[-0.9, 0.9]$  &  $[-0.8, 0.9]$  & 1.8  \\
\hline
\end{tabular}
\label{tab:VBS_aQGC}
\end{table}

\section{Summary}
\label{sec:summary}
The cross section for the electroweak production of a \PW~boson, a photon, and two jets is measured in proton-proton collisions at a center-of-mass energy of 13\TeV. 
The data correspond to an integrated luminosity of 35.9\fbinv collected with the CMS detector. 
Events are selected by requiring one identified lepton (electron or muon), a moderate missing transverse momentum, one photon, and two jets with a large rapidity separation and a large dijet mass.
The observed significance is 4.9 standard deviations, where a significance of 4.6 standard deviations is expected based on the standard model. 
After combination with previously reported CMS results based on 8\TeV data, the observed (expected) signal significance is 5.3 (4.8) standard deviations. 
This constitutes the first observation of electroweak {\Wgjj} production in proton-proton collisions. 
The cross section for the electroweak {\Wgjj} production in a restricted fiducial region is measured as $20.4\pm 4.5\unit{fb}$ and the total cross section for \Wg production in association with 2 jets in the same fiducial region is $108\pm 16\unit{fb}$, consistent with standard model predictions. 
Constraints placed on anomalous quartic gauge couplings in terms of dimension-8 effective field theory operators are competitive with previous results. 
For the parameters \fcoefflam{M}{2--5} and \fcoefflam{T}{6--7}, the constraints are the most stringent to date.

\begin{acknowledgments}
  We congratulate our colleagues in the CERN accelerator departments for the excellent performance of the LHC and thank the technical and administrative staffs at CERN and at other CMS institutes for their contributions to the success of the CMS effort. In addition, we gratefully acknowledge the computing centers and personnel of the Worldwide LHC Computing Grid for delivering so effectively the computing infrastructure essential to our analyses. Finally, we acknowledge the enduring support for the construction and operation of the LHC and the CMS detector provided by the following funding agencies: BMBWF and FWF (Austria); FNRS and FWO (Belgium); CNPq, CAPES, FAPERJ, FAPERGS, and FAPESP (Brazil); MES (Bulgaria); CERN; CAS, MoST, and NSFC (China); COLCIENCIAS (Colombia); MSES and CSF (Croatia); RIF (Cyprus); SENESCYT (Ecuador); MoER, ERC IUT, PUT and ERDF (Estonia); Academy of Finland, MEC, and HIP (Finland); CEA and CNRS/IN2P3 (France); BMBF, DFG, and HGF (Germany); GSRT (Greece); NKFIA (Hungary); DAE and DST (India); IPM (Iran); SFI (Ireland); INFN (Italy); MSIP and NRF (Republic of Korea); MES (Latvia); LAS (Lithuania); MOE and UM (Malaysia); BUAP, CINVESTAV, CONACYT, LNS, SEP, and UASLP-FAI (Mexico); MOS (Montenegro); MBIE (New Zealand); PAEC (Pakistan); MSHE and NSC (Poland); FCT (Portugal); JINR (Dubna); MON, RosAtom, RAS, RFBR, and NRC KI (Russia); MESTD (Serbia); SEIDI, CPAN, PCTI, and FEDER (Spain); MOSTR (Sri Lanka); Swiss Funding Agencies (Switzerland); MST (Taipei); ThEPCenter, IPST, STAR, and NSTDA (Thailand); TUBITAK and TAEK (Turkey); NASU (Ukraine); STFC (United Kingdom); DOE and NSF (USA).
   
  \hyphenation{Rachada-pisek} Individuals have received support from the Marie-Curie program and the European Research Council and Horizon 2020 Grant, contract Nos.\ 675440, 752730, and 765710 (European Union); the Leventis Foundation; the A.P.\ Sloan Foundation; the Alexander von Humboldt Foundation; the Belgian Federal Science Policy Office; the Fonds pour la Formation \`a la Recherche dans l'Industrie et dans l'Agriculture (FRIA-Belgium); the Agentschap voor Innovatie door Wetenschap en Technologie (IWT-Belgium); the F.R.S.-FNRS and FWO (Belgium) under the ``Excellence of Science -- EOS" -- be.h project n.\ 30820817; the Beijing Municipal Science \& Technology Commission, No. Z191100007219010; the Ministry of Education, Youth and Sports (MEYS) of the Czech Republic; the Deutsche Forschungsgemeinschaft (DFG) under Germany's Excellence Strategy -- EXC 2121 ``Quantum Universe" -- 390833306; the Lend\"ulet (``Momentum") Program and the J\'anos Bolyai Research Scholarship of the Hungarian Academy of Sciences, the New National Excellence Program \'UNKP, the NKFIA research grants 123842, 123959, 124845, 124850, 125105, 128713, 128786, and 129058 (Hungary); the Council of Science and Industrial Research, India; the HOMING PLUS program of the Foundation for Polish Science, cofinanced from European Union, Regional Development Fund, the Mobility Plus program of the Ministry of Science and Higher Education, the National Science Center (Poland), contracts Harmonia 2014/14/M/ST2/00428, Opus 2014/13/B/ST2/02543, 2014/15/B/ST2/03998, and 2015/19/B/ST2/02861, Sonata-bis 2012/07/E/ST2/01406; the National Priorities Research Program by Qatar National Research Fund; the Ministry of Science and Higher Education, project no. 02.a03.21.0005 (Russia); the Programa Estatal de Fomento de la Investigaci{\'o}n Cient{\'i}fica y T{\'e}cnica de Excelencia Mar\'{\i}a de Maeztu, grant MDM-2015-0509 and the Programa Severo Ochoa del Principado de Asturias; the Thalis and Aristeia programs cofinanced by EU-ESF and the Greek NSRF; the Rachadapisek Sompot Fund for Postdoctoral Fellowship, Chulalongkorn University and the Chulalongkorn Academic into Its 2nd Century Project Advancement Project (Thailand); the Kavli Foundation; the Nvidia Corporation; the SuperMicro Corporation; the Welch Foundation, contract C-1845; and the Weston Havens Foundation (USA).
\end{acknowledgments}

\bibliography{auto_generated}
\cleardoublepage \appendix\section{The CMS Collaboration \label{app:collab}}\begin{sloppypar}\hyphenpenalty=5000\widowpenalty=500\clubpenalty=5000\vskip\cmsinstskip
\textbf{Yerevan Physics Institute, Yerevan, Armenia}\\*[0pt]
A.M.~Sirunyan$^{\textrm{\dag}}$, A.~Tumasyan
\vskip\cmsinstskip
\textbf{Institut f\"{u}r Hochenergiephysik, Wien, Austria}\\*[0pt]
W.~Adam, F.~Ambrogi, T.~Bergauer, M.~Dragicevic, J.~Er\"{o}, A.~Escalante~Del~Valle, R.~Fr\"{u}hwirth\cmsAuthorMark{1}, M.~Jeitler\cmsAuthorMark{1}, N.~Krammer, L.~Lechner, D.~Liko, T.~Madlener, I.~Mikulec, F.M.~Pitters, N.~Rad, J.~Schieck\cmsAuthorMark{1}, R.~Sch\"{o}fbeck, M.~Spanring, S.~Templ, W.~Waltenberger, C.-E.~Wulz\cmsAuthorMark{1}, M.~Zarucki
% ...
\vskip\cmsinstskip
\dag: Deceased\\
1:  Also at Vienna University of Technology, Vienna, Austria\\
% ...
\end{sloppypar}
\end{document}